\documentstyle[12pt,aasms4]{article} 
\tightenlines 
 
\newcommand\be{\begin{equation}} 
\newcommand\ee{\end{equation}} 
 
\input epsf 
\lefthead{Basu et al.} 
\righthead{The Effect of reference models on helioseismological results} 
\begin{document} 
 
\title{\bf How much do helioseismological inferences depend upon the 
assumed reference model?} 
 
\author{Sarbani Basu} 
\affil{Institute for Advanced Study, Olden Lane, Princeton, NJ 08540, U. S. 
A.} 
\author{M. H. Pinsonneault} 
\affil{Department of Astronomy, Ohio State University, Columbus, Ohio
43210, U. S. A.} 
\author{John N. Bahcall} 
\affil{Institute for Advanced Study, Olden Lane, Princeton, NJ 08540, U. S. 
A.}

\begin{abstract} 
We investigate systematic uncertainties in determining 
 the profiles of the solar sound speed, 
density, and adiabatic index  by
helioseismological techniques. 
We find that  rms uncertainties ---  averaged over the sun  --- of
$\sim 0.02\%-0.04$\% are contributed
to the sound speed profile by each of three sources: 1) the choice of
assumed reference model, 2) the width of the inversion kernel, and 3)
the measurement errors.  The rms agreement between the standard solar
model sound speeds and the best helioseismological determinations is about
$0.07$\%. 
The profile of the 
adiabatic index $\Gamma_1$
is determined to an accuracy of about $0.02$\% with the MDI data set.
The density profile is about an order of magnitude less well
determined by the helioseismological measurements. 
Five state-of-the-art models, each with a significant difference in
 the input physics or a parameter choice, 
all give comparably good agreement with global
 helioseismological measurements.
We consider four deficient solar models that are constructed using either old
 input data, assuming the $^3$He + $^4$He fusion reaction does not
 occur, neglecting element diffusion, or  artifically 
mixing the interior of the sun.
When used as reference models in the inversion process, 
these deficient  models yield sound speeds for the sun that differ
 only by $0.1$\% from the sound speeds obtained using the standard model.
We conclude that 
even relatively crude reference models yield reasonably accurate solar
 parameters. 
Although acceptable for most purposes as reference models,
non-standard solar models in which the core is artifically mixed or
 in which element diffusion is neglected are strongly disfavored by 
 the p-mode oscillation data.
These non-standard  models produce sound speed profiles with
 respect to the sun that are, respectively, $4.5$ and $18$ times worse
 than the agreement obtained with the standard solar model.
\end{abstract} 
 
\keywords{Sun: oscillations; Sun: interior} 
 
\section{Introduction} 
 \label{sec:introduction}
 
Helioseismology has revolutionized our knowledge of the Sun and
enriched, stimulated,  and largely 
validated our understanding of the  evolution of main sequence stars 
with masses comparable to the sun.
The statistical and measurement errors in the best modern 
samples of helioseismological frequencies have been reduced 
to tiny fractional values, $\sim$ a few times $10^{-5}$, which lead to
fractional errors in, e. g., the inferred sound velocities that are
formally of order
$10^{-4}$ or less.

We therefore concentrate in this paper on systematic uncertainties.
The main result of this paper is that the systematic uncertainties in
determining the sound speeds are about an order of magnitude larger
than the statistical errors.

We need to know quantitatively the accuracy of the   
solar sound speed and density profiles that are inferred from
helioseismology since these characteristics are often used, e. g.,  to 
test opacity calculations (Korzennik and Ulrich 1989,
Tripathy, Basu \& Christensen-Dalsgaard 1997),
to investigate equations 
of state (Ulrich, 1982; Elliott \& Kosovichev 1998;
Basu, D\"appen \& Nayfonov 1999),
and to derive abundance profiles in the Sun
(Antia \& Chitre 1998; Takata \& Shibahashi 1998).
Moreover, the precise agreement between the inferred sound speeds and
those calculated with a standard solar model used to predict solar
neutrino fluxes is a strong argument that solar neutrino
experiments require new physics, not revised astrophysics (Bahcall,
et al.  1997; Bahcall, Basu, \& Pinsonneault 1998).

Solar oscillation data have been inverted to determine the 
sound speed and density profiles over nearly the entire  Sun 
(cf. Dziembowski, Pamyatnykh \& 
Sienkiewicz 1990; D\"appen et al. 1991; Antia \& Basu 1994; 
 Kosovichev et al. 1997), as well as the 
adiabatic index $\Gamma_1$ 
(cf. Antia \& Basu 1994; Elliott 1996; Elliott \& Kosovichev 1998). 
However,  helioseismological determinations of  solar parameters 
generally proceed through a linearization 
of the equations of stellar oscillations around a theoretical  reference 
model of the sun. 
We are therefore naturally led to ask the question which is reflected
in the
title of this paper: How dependent are the inferred solar
characteristics upon the assumed reference model?

Here is how we have answered this question.
We have  constructed a broad range
of conceivable modern solar models,  each model with 
different input physics or assumptions, and used this broad set of
models as reference models to calculate solar sound speeds and
densities with different data sets.
The range of inferred sound speeds and densities define empirically the
systematic uncertainties that are inherent in using references models
to invert helioseismological data.

Five of the solar models are roughly comparable to each other and use
physics that is within the currently acceptable range; these models
are defined in  \S~\ref{sec:models}, where they are given the names
STD, PMS,ROT,R78,R508.  We also use four different models, each of
which is deficient in one or more major aspects of the input
physics. The deficient models are also defined in \S~\ref{sec:models},
where they are named OLD, S$_{34} = 0$, NODIF, and MIX.

To explore  the systematic uncertainties associated with choosing a
particular data set, we have chosen three different sources of
observational data.
We have used solar frequencies 
obtained by: 1) the Michelson Doppler Imager (MDI) instrument on board 
the Solar and Heliospheric Observeratory (SOHO) during the first 144 days of 
its 
operation (cf. Rhodes et al. 1997);  2) the set of frequencies 
obtained by observations of the Global Oscillation Network Group
(GONG) between  months $4$--$14$  of its observations; 3) a 
combination of the data 
from the Birmingham Solar Oscillation Network 
(BiSON; cf. Chaplin et al. 1996)  
and the  Low-$\ell$ (LOWL; cf. Tomczyk et al. 1995) 
instrument. The third set is  same as the one used by Basu et al. (1997) and 
is described in detail there.  
The MDI and GONG sets have 
good coverage of intermediate degree modes. The MDI set has p-modes 
from $\ell = 0$
up to a degree of $\ell=194$ while the GONG set has modes 
from $\ell = 0$
up to $\ell=150$.  
However, both these sets  are somewhat deficient in low degree modes. 
The BiSON+LOWL combination, on the other hand, has a better coverage of 
low degree modes, but has modes from $\l = 0$ only up to $\ell=99$. 

We concentrate in this paper on the properties of helioseismological
inversions. We discuss the implications for neutrino physics in 
Bahcall, Pinsonneault \& Basu (1999, in preparation) and explore the
results of mixing and rotation on element abundances and
helioseismology in 
Pinsonneault, Basu \& Bahcall (2000, in preparation).
 
The present paper is organized as follows. We summarize briefly the inversion
technique in \S~\ref{sec:technique} and then describe the solar 
models used in
\S~\ref{sec:models}.  Before investigating the systematic
uncertainties due to the choice of reference model, we first
investigate in \S~\ref{sec:datadiffs} the uncertainties due to the
choice of data set and in \S~\ref{sec:resolution} the uncertainties due
to the finite resolution of the inversion kernel. We present the
inferred solar profiles of sound speed and density in 
\S~\ref{sec:stdsounddensity} and Table~\ref{tab:solardata}. 
We compare in \S~\ref{sec:comparestd}
the standard solar model with the helioseismological
measurements  and 
compare in 
\S~\ref{sec:compvariants} the eight variant models with the observations .
The dependence of the profiles of the sound speed,  the density, and the
adiabatic index  upon
the assumed reference model is determined in
\S~\ref{sec:dependence}.
We summarize and discuss our principal results in 
\S~\ref{sec:discussion} .

\section{Inversion technique} 
\label{sec:technique} 

The equations describing linear and adiabatic stellar oscillations 
are known to be Hermitian (cf. Chandrasekhar 1964). This property 
of the equations can be used to relate the 
differences  between the 
structure of the Sun and the reference model  to the 
differences in the frequencies of the Sun and the model by known kernels. 
Non-adiabatic effects and other errors in modeling the surface layers 
give rise to frequency shifts (Cox \& Kidman 
1984; Balmforth 1992) that are not accounted for by the variational 
principle. In the absence of a more fundamental method, these surface
effects have been treated by the   {\it ad hoc} procedure of 
including an arbitrary function of frequency in the variational 
formulation. 
 
When the oscillation equation is linearized --- under the assumption 
of hydrostatic equilibrium --- the fractional change in the 
frequency can be related to the fractional changes in the squared 
sound-speed ($c^2$) and density ($\rho$). 

The sound-speed 
$c$ used here is the adiabatic sound speed, which is defined  as
\be
c^2={\Gamma_1 P\over \rho},
\label{eqn:sound}
\ee
where $\Gamma_1$,  the adiabatic index, is defined by the
thermodynamic relation 
\be
\Gamma_1=\left({\partial \ln P\over\partial \ln \rho}\right)_s,
\label{eqn:gamdef}
\ee
Here $P$ is the pressure and $s$ is 
the entropy.

We can write,
\begin{equation} 
\frac{\delta\omega_i}{\omega_i}  \! = \!  
\int K_{c^2,\rho}^i (r) \frac{\delta 
c^2}{c^2}(r) d r 
+ \int K_{\rho,c^2}^i (r) \frac{\delta \rho}{\rho}(r) d r  
 + \frac{F_{\rm surf}(\omega_i)}{E_i}\; , 
\label{eqn:freqdif} 
\end{equation} 
(cf, Dziembowski et al. 1990).  Here $\delta \omega_i$ is the 
difference in the frequency $\omega_i$ of the $i$th mode between the 
solar data and a reference model.  
The kernels $K_{c^2,\rho}^i$ 
and $K_{\rho,c^2}^i$ are known functions of the reference model which 
relate the changes in frequency to the changes in $c^2$ and $\rho$,
respectively; and $E_i$ is the inertia of the mode, normalized by the 
photospheric amplitude of the displacement.  The term $F_{\rm surf}$ 
results from the near-surface errors. 
 
 In this work we have  used the subtractive optimally localized 
averages (SOLA) method (cf, Pijpers \& Thompson 1992) to invert 
Eq.~(\ref{eqn:freqdif}) in order to determine  
$\delta c^2/c^2$ and $\delta\rho/\rho$ between a reference model and 
the Sun.  The principle of the inversion technique 
is to form linear combinations of Eq.~(\ref{eqn:freqdif}) with  
weights $d_i(r_0)$ chosen so as to obtain an average  of 
$\delta c^2/c^2$ (or $\delta\rho/\rho$) localized near 
$r=r_0$, while suppressing the contributions from 
$\delta\rho/\rho$ when inverting for $\delta c^2/c^2$  (or $\delta c^2/c^2$ 
when inverting for $\delta\rho/\rho$ ) and the near-surface errors. 
In addition, the statistical errors in the combination must be 
constrained. We define the {\it averaging kernel} as 
\be 
{\cal K}(r_0,r)=\sum d_i(r_0)K^i_{c^2,\rho}(r), 
\label{eqn:avker} 
\ee 
which is normalized so that $\int {\cal K}(r_0,r)dr =1$.
The results of the inversion determine, for example, the difference,
$\delta c^2$ between the square of the model sound speed
and the square of the sound speed of the sun, i. e.,
\be
\delta c^2 = c^2_{\rm
\odot} -c^2_{\rm model} .
\label{eqn:deltac^2}
\ee
If the inversion is successful, the relative sound-speed difference 
can be written as
\be 
\left( {\delta c^2\over c^2}\right )_{\rm inv}(r_0)\equiv\int 
{\cal K}(r_0,r){\delta c^2\over c^2}dr  
\simeq \sum  d_i(r_0){\delta\omega_i\over\omega_i}. 
\label{eqn:inv} 
\ee 
The  second expression is only approximately equal to the third expression in 
Eq.~(\ref{eqn:inv})  because 
contributions from the second and third terms in Eq.~(\ref{eqn:freqdif}) 
cannot be eliminated completely. 
The weights $d_i$ are determined such that 
these contributions are substantially less than the error in the solution 
because of measurement errors in the data. 
An  expression similar to Eq.~(\ref{eqn:inv}) can be written  
for $\left(\delta\rho/\rho\right )_{\rm inv}(r_0)$ as well. 

The averaging kernels, $\cal K$, determine the extent to which we obtain a  
localized average of ${\delta c^2/ c^2}$ or  
$\delta\rho/\rho$. The width of the averaging kernel, e.g., 
the distance between the first and third quartile points, provides 
a measure of the resolution. Ideally, one would like the averaging kernel 
to be a $\delta$-function at $r=r_0$, but since only a finite amount of data 
is available, that is impossible to achieve. 
The effect of finite resolution 
on the inferred values of $\delta c^2/c^2$ was studied by 
Bahcall, Basu \& Kumar (1997) and  found to be small
for contemporary data sets.

The errors in the inversion results are calculated assuming that the
errors in the frequencies are uncorrelated. Thus the error in
$(\delta c^2/c^2)_{\rm inv}$ at radius $r_0$ is given by
\be
\sigma^2(r_0)=\sum_i d^2_i(r_0){\epsilon_i\over\omega_i}^2,
\label{eqn:error}
\ee
where, $\epsilon_i$ is the quoted error of mode $i$ with
frequency $\omega_i$.

The adiabatic index $\Gamma_1$ is related to the sound speed
and density [cf., Eq.~\ref{eqn:sound}]. Hence, 
the kernels of $c^2$ and $\rho$
can be easily converted to those of $\Gamma_1$ and $\rho$
(and vice versa, see e.g., Gough 1993).

The details of how the method is implemented  can be found in  
Basu et al. (1996) and the effect of the inversion parameters on the
results  are 
discussed in Rabello-Soares, Basu \& Christensen-Dalsgaard  (1999).

\section{Models used} 
\label{sec:models}
 
\begin{figure} 
\plotone{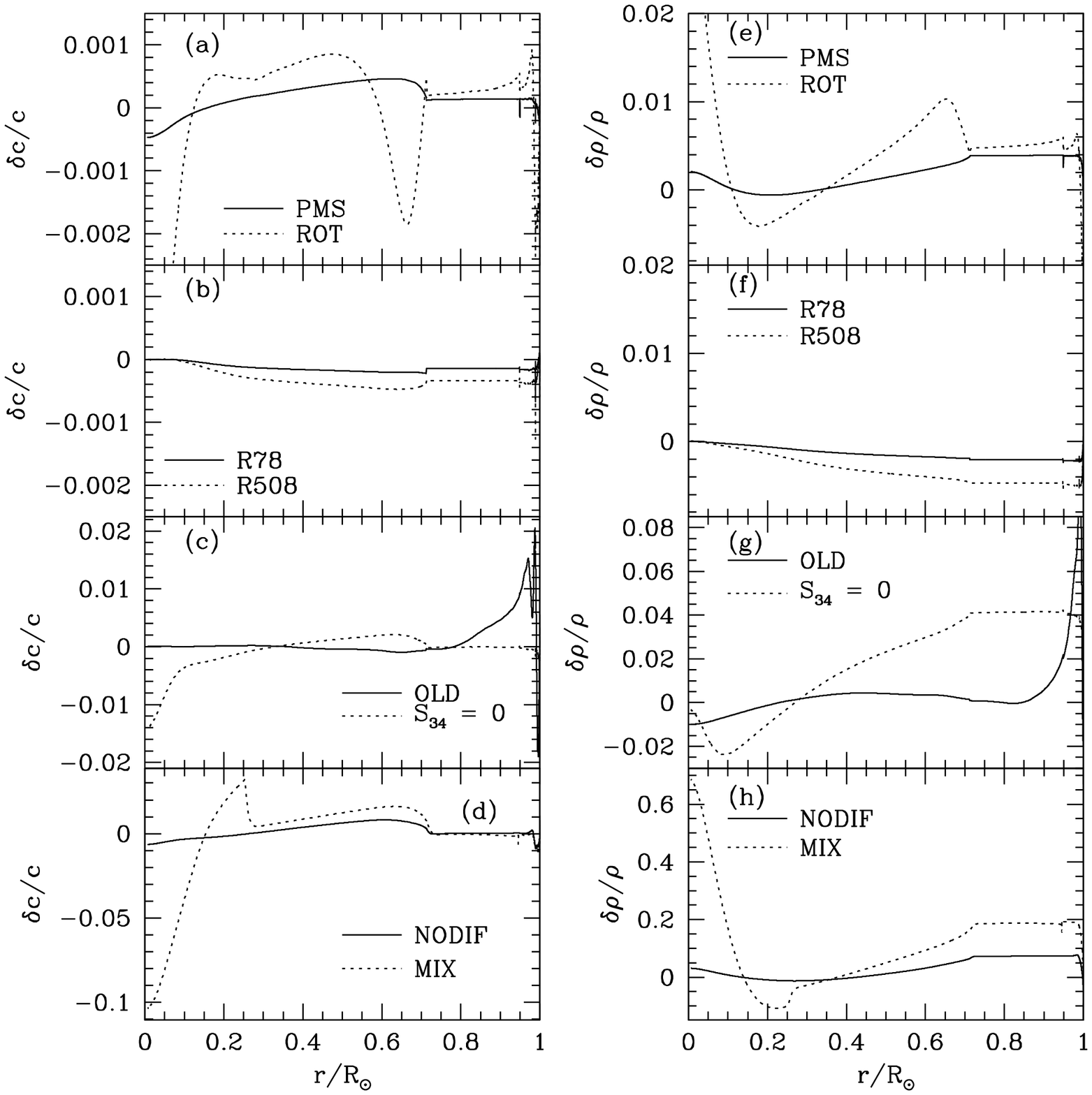} 
\figcaption{ The relative sound-speed differences, $\delta c/c$, and the 
relative density differences, $\delta\rho/\rho$, between the standard model 
STD 
and the other solar models. For each model, 
 $\delta c/c=(c_{\rm STD}-c_{\rm model})/c_{\rm model}$ and 
$\delta\rho/\rho=(\rho_{\rm STD}-\rho_{\rm model})/\rho_{\rm model}$. 
The models are described in Table~\ref{tab:diffmodels}.
The vertical scales for panels c), d), g), and h) are larger than for
the other panels.
\label{fig:diffmodels}
 } 
\end{figure}

We have used a set of nine solar models as reference models to invert  
the three sets of helioseismological data.
Figure~\ref{fig:diffmodels} compares the computed 
sound speeds and densities of eight
variant and deficient solar  
models discussed in \S~\ref{subsec:variants} and
\S~\ref{subsec:deficient}
with the sound
speeds and densities 
computed for our standard model, STD, which is described in
\S~\ref{subsec:std}.

Table~\ref{tab:diffmodels} summarizes some of the key properties of
the solar models discussed in this section. 
The convective zone (CZ) for most of the 
 models are very  close to the observed value
of $0.713\pm 0.001$ (cf, Basu \& Antia, 1997). The only models for
which the depth of the convective zone is clearly  wrong are the
NODIF and MIX  models. 
The surface helium abundance of most of the models  
is also consistent with the
abundance determined helioseismologically, $0.248 \pm 0.003$ (Basu 1998). 
Only the 
 NODIF model has a helium abundance that is obviously inconsistent with the
observed helium abundance.
 
\subsection{Standard Model: STD}
\label{subsec:std}

Our standard model, STD,  is
constructed with the OPAL 
equation of state (Rogers, Swenson \& Iglesias 1996) 
and OPAL opacities 
(Iglesias \& Rogers 1996) which are supplemented by the low 
temperature opacities of Alexander \& Ferguson (1994). The model was 
constructed using the 
usual mixing length formalism for calculating convective flux. 
The nuclear reaction rates were calculated with the subroutine 
exportenergy.f (cf. Bahcall and Pinsonneault 1992), using the reaction
data in Adelberger et al. (1998) and with electron and ion 
weak screening as indicated by recent calculations of 
Gruzinov \& Bahcall (1998; see also Salpeter,  1954). 
The model 
incorporates helium and heavy element diffusion 
using the exportable diffusion subroutine of Thoule (cf. Thoule, 
Bahcall \& Loeb, 1994; Bahcall \& Pinsonneault 1995)\footnote{Both the
nuclear energy generation subroutine, exportenergy.f, and the
diffusion subroutine, diffusion.f, 
are available at the Web site www.sns.ias.edu/$~$jnb/,menu item:
neutrino software and data.}.

For the standard model, the evolutionary calculations were
started at the main-sequence stage. 
The model has a radius of 695.98 Mm.
The 
ratio of heavy elements to hydrogen ($Z/X$) at the surface of the model is 
0.0246, which is consistent with the value observed by  
Grevesse \& Noels (1993). A  Krishnaswamy $T$-$\tau$ relationship for
the atmosphere was used. 

Earlier inversions with similar models have shown that the 
difference in sound-speed between standard solar 
models and the Sun is 
small, of the order of $0.1$\% rms (cf. Bahcall et al. 1997; 
Basu et al. 1997; Bahcall, Basu \& Pinsonneault  1998).

\subsection{Variant models}
\label{subsec:variants}
 
In this subsection, we describe four models that are slight variations
on the theme of the standard model.

Model PMS is evolved from the 
pre-main sequence stage, but otherwise is the same as STD.
The difference in internal structure 
that results from  including the 
pre-main sequence evolution is known to be very 
small (Bahcall \& Glasner 1994).

Model ROT incorporates mixing induced by rotation and is a reasonable
upper bound to the degree of rotational mixing which is consistent
with the observed depletion of lithium in the sun
(cf. Pinsonneault 1996, Pinsonneault et al. 1999). 
The initial rotation period for the model is 8 days, which is the
median observed value for T Tauri stars (Choi \& Herbst 1996).
The structural effects of rotation were treated using the
method of Kippenhahn \&  Thomas (1970) 
  Rigid rotation as a function of
depth was enforced at all times in convective regions; in radiative
regions the transport of angular momentum and the associated mixing
were solved with coupled diffusion equations (see section 5  in
Pinsonneault 1997).  Angular momentum loss from a
magnetic wind is included, and then the thermal structure and angular
momentum distribution in the interior are used to determine the diffusion
coefficients as a function of depth.  The angular momentum loss rate and
the velocity estimates for the diffusion coefficients are
the same as Krishnamurthi et al.  (1997).
The parameters of the model, which is started in the pre-main sequence
phase,  are fixed as follows:
(1) the model is required to reproduce the 
equatorial surface rotational  period of $25.4$ days;  and (2) the
model is required to 
reproduce the observed solar lithium depletion of
$2.19$ dex (the difference between the meteoritic Li abundance
of $3.34$ on the logarithmic scale where H=12 and the photospheric
Li abundance of $1.15$). The rotational model neglects
angular momentum transport by internal magnetic fields and gravity
waves; both of these mechanisms can transport angular momentum
without mixing and therefore  reduce the angular momentum content in
the core,  decreasing  the mixing from meridional circulation and
different instabilities.  There is  evidence 
from helioseismology that additional angular momentum transport
mechanisms, such as gravity waves or magnetic fields, are needed to
explain the absence of strong differential rotation with depth in the
solar core (see Tomczyk, Schou \& Thompson 1995).

There has been considerable discussion recently regarding the precise
value of the solar radius 
(cf. Antia 1998; Schou et al. 1997;  Brown \& Christensen-Dalsgaard 1998) 
and some discussion of the 
effects of the uncertainty in radius on the quantities inferred from
the helioseismological inversions (cf. Basu 
1998),  We have therefore considered
two models which were constructed with the same input physics as STD, 
but which have model radii which differ from the radius assumed in
constructing STD. 

Model R78 has  a radius of 695.78 Mm, 
which is the radius that has been  determined from the frequencies of f-modes 
(cf. Antia 1998). Model R508 has a radius of 695.508 Mm,  which is the 
solar radius as determined 
by Brown \& Christensen-Dalsgaard (1998), who used the measured
duration of solar meridian transits during the 6 years 1981--1987 and
combined these measurements with models of the solar 
limb-darkening function to estimate the value of the solar radius.
The solar structure is affected only very slightly by the choice of
model radii.  The fractional differences in the model radii considered
in this paper are less than $1$ part in $10^3$, whereas the radial resolution
in the sound speed is at best a few percent (see Figure 1 of Bahcall,
Basu \& Kumar 19u97).

The rms sound speed differences between the variant models and the STD
model are: $0.03$\% (PMS), $0.08$\% (ROT), $0.15$\% (R78), and  $0.03$\%
(R508).  The average  difference (rms) between the four variant models
and the STD model is $0.07$\% .

\subsection{Deficient models}
\label{subsec:deficient}

In this subsection, we describe four models that are each deficient in
one or more significant aspects of the input physics.

The model, OLD, is a standard solar model  constructed with
some relatively old physics: 
the Yale equation of state (cf. Guenther et al. 1992) with the 
Debye-H\"uckel correction (cf. Bahcall, Bahcall \& Shaviv 1968) and old OPAL 
opacities (Iglesias, Rogers \& Wilson 1992) supplemented with 
low temperature opacities from Kurucz (1991). The model does include
helium and heavy element diffusion and uses the nuclear reaction cross
section factors ($S_0$) from Adelberger et al. (1998). In the course
of writing this paper, we uncovered a small inconsistency in the code
for the Yale equation of state.  Fortunately, this inconsistency(which
was introduced in recent revisions) does not affect any of our
published results which no longer use the Yale equation of state. For
the OLD model, the error in the code causes an 
increase in the mean
molecular weight at fixed composition of 0.1\% relative to the correct value.

The OLD model differs from the STD model in using a cruder equation of
state and less precise radiative
opacities.
Using the old physics rather than the current best input data as was
done in constructing
STD causes significant changes, primarily in the convection
zone. This is a typical signature for large differences in the input
equation of state, which is the most significant physics deficiency of
this model.

For Model S$_{34} = 0$,  the cross-section of the nuclear 
reaction $^3$He($\alpha$,$\gamma$)$^7$Be was set  equal to $0$ 
in order to minimize the  
calculated neutrino capture rates in the Gallex and SAGE experiments 
(see Bahcall 1989, Chapter 11). This assumption contradicts many
laboratory experiments  which measured a  cross section  for the 
$^3$He($\alpha$,$\gamma$)$^7$Be that is competitive with the other way
of terminating the $p-p$ chain, namely, $^3$He($^3$He,2p)$^4$He .
For the S$_{34} = 0$ model, nuclear
fusion energy is achieved in a significantly
different way than for the standard solar model and therefore the
calculated solar structure is appreciably  different from the
standard model (Bahcall and Ulrich 1988). In the standard solar model,
about $15$\% of the terminations of the $p-p$ chain involve the
$^3$He$(\alpha,\gamma)^7$Be reaction, whose rate is proportional to
S$_{34}$. If we artificially choose S$_{34} = 0$, then in this model
$^7$Be is not produced and there are no $^7$Be or ${}^8$B neutrinos. 
 
Model NODIF does not  include either helium or heavy-element
 diffusion.
This model 
therefore represents
the state-of-the art in solar modeling  prior to 1992 (cf. 
Bahcall \& Ulrich 1988, Bahcall \& Pinsonneault 1992, Proffitt 1994).

Model MIX has an artificially 
 mixed core. The inner $50$\% by mass ($25$\% by radius) was required
to be chemically homogeneous at all times.  All of the other ingredients
of the model, including helium and heavy element diffusion, are the
same as in model STD.
This model was constructed to be similar to 
the prescription of Cumming \& Haxton (1996), who changed  by hand 
the $^3$He abundance as a function
of radius in the final BP95 (Bahcall \& Pinsonneault 1995) 
solar model in order to minimize the discrepancy between measurements
of the total event rates in neutrino experiments and the calculated
event rates. Since the sun evolves over time and Cumming and Haxton
 only changed the abundances in the final model, we had to adopt some
 additional 
prescription as to how the mixing proceeds as a function of time. We
 assumed that the mixing was infinitely effective (the core was fully
 mixed) and constant in time.

The rms sound speed differences between the deficient models and the STD
model are: $0.4$\% (NODIF), $0.2$\% (OLD), $0.2$\% ($S_{34} = 0$), and  
$1.9$\%(MIX).  The average rms difference between the deficient models
and the STD model is $0.7$\% , which is an order of magnitude larger
than for the variant models discussed in \S~\ref{subsec:variants}.
 
\section{How accurate are the measurements?}
\label{sec:datadiffs}

How similar are the sound speeds, densities, and values of $\Gamma_1$
 inferred from different
data sets? 
For sound speeds, this question is answered in
Fig.~\ref{fig:diffdata}(a)-(c), which  shows 
the sound-speed  
difference between the  standard model STD and the Sun as obtained
using the MDI, GONG, and BiSON+LOWL data sets. 
The results appear relatively similar to the eye, but there are some
differences as large as $0.05$\% inside $0.2 R_\odot$. 
For densities and $\Gamma_1$, the differences between the standard
 model and the Sun  are
 illustrated in Fig.~\ref{fig:diffdatagam}. 

\subsection{The sound speed}
\label{subsec:datasoundspeeds}

\begin{figure} 
\plotfiddle{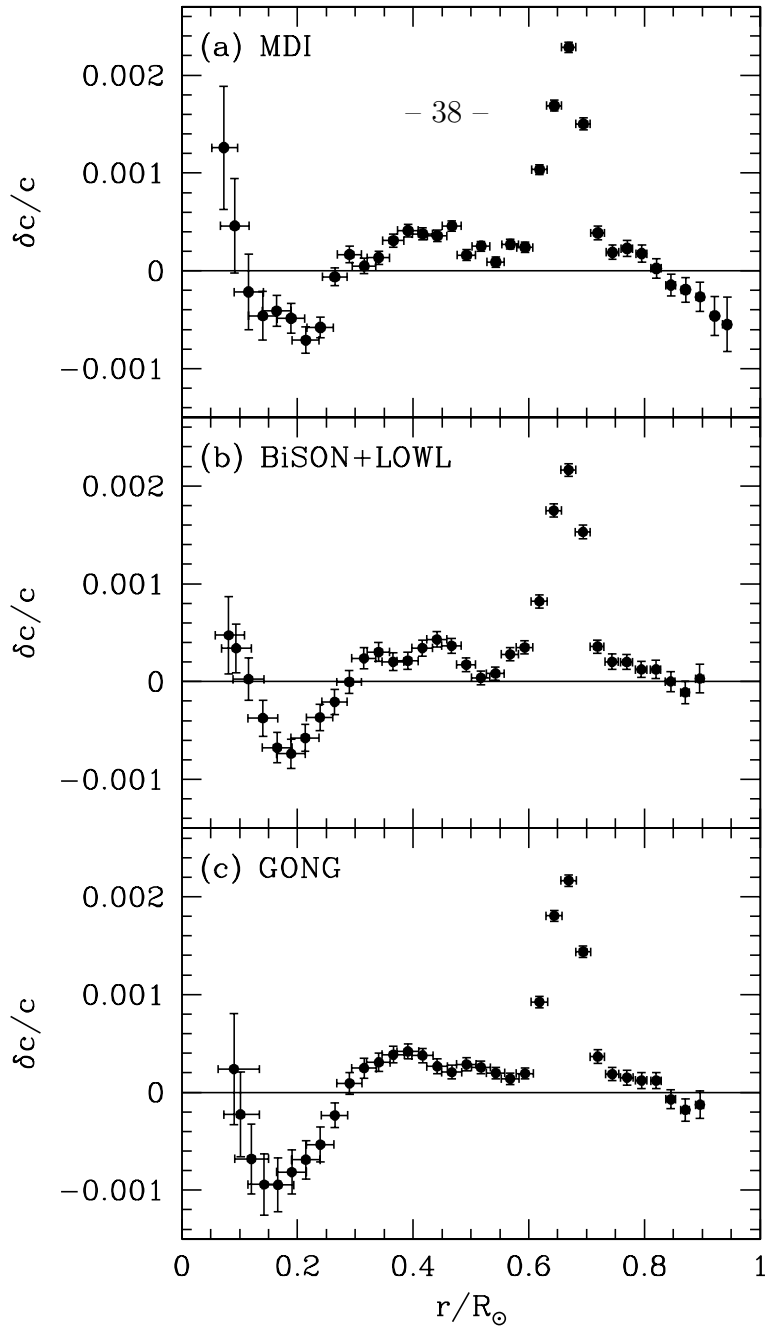}{13 true cm}{0}{90}{90}{-280}{-130}
\figcaption{ The relative sound-speed difference between the Sun and 
model STD obtained using three different solar oscillation data-sets 
[Panels (a),(b) and (c)].
The differences are in the sense (Sun $-$ Model)/Model. 
The vertical error-bars indicate $1\sigma$ errors in the inversion  
results because of errors in the data. The horizontal error-bars 
are a measure of the resolution of the inversion.
\label{fig:diffdata} } 
\end{figure}

Figure~\ref{fig:sumdiffdata}a and Figure~\ref{fig:sumdiffdata}b
reveal even  more clearly the differences between the three data sets.
In Fig.~\ref{fig:sumdiffdata}  we plot
the difference in sound speeds obtained with the 
GONG and the MDI data and the difference in sound speeds obtained with
BiSON+LOWL and MDI.
In all the panels shown in Fig.~\ref{fig:diffdata} and
Fig.~\ref{fig:sumdiffdata}, only one solar model, STD, has been used. 
The resolution of the inversions obtained with the three data sets is
almost the same, hence the errors due to resolution should be 
very similar in each set. The only exception occurs near  the solar 
surface, where the fact that 
the three  data sets have different high-degree coverage becomes important.
The extent of the high-degree coverage
is probably the cause of the systematic differences in the
sound speeds that are seen in Fig.~\ref{fig:diffdata} and
Fig.~\ref{fig:sumdiffdata} near the surface. Elsewhere, 
differences are caused solely by the measurements.

The errors in the velocity measurements are apparently reasonably well
understood; nearly all of the points lie within the $2\sigma$ error
bounds delineated in Fig.~\ref{fig:sumdiffdata}. If there were
large  systematic errors in one of the experiments, then we would
have expected to see values for $\Delta c/c$ in Fig.~\ref{fig:sumdiffdata}
that fell outside the $2\sigma$ limits.
The errors shown in Fig.~\ref{fig:sumdiffdata} were calculated by
combining quadratically at each target radius the errors obtained
for inversions of   the MDI set and the other sets.
The errors at each  radius were evaluated as per Eq.~\ref{eqn:error}.

All data sets yield results for the sound speed profile 
that are consistent with each other within  the errors of the data sets
(cf. ~Fig.~\ref{fig:sumdiffdata}).
The rms differences are only $0.02$\% for the 
sound speeds calculated with  the BiSON + LOWL and the MDI data sets and also
$0.02$\% for the differences found between the sound speeds calculated
with the GONG and the MDI data sets.

\subsection{The density profile}
\label{subsec:datadensityprofile}

The density profile cannot be determined as precisely as the profile
of the sound speed.
The primary reason for the reduced precision in inverting for the
density profile is that the $p-$mode oscillation  frequencies are determined
predominantly by the sound speed, with the density contributing
only through the perturbation of the gravitational potential.
In fact in the asymptotic limit, the frequencies are
determined by sound-speed alone. As a result there is relatively 
little information about density in the frequencies.
A further difficulty that must be overcome in a density inversion is 
the precise satisfaction of the condition for mass conservation, 
\be
\int_0^R 4\pi r^2 \delta \rho = 0.
\label{eqn:masscon}
\ee
In order to satisfy this condition with high numerical accuracy, 
the density must be reasonably well
determined at  all radii. Equation~(\ref{eqn:masscon}) therefore 
requires a  set of oscillation frequencies that includes a large number of
low degree modes (to invert accurately for the core ) as well as a
large number of high degree modes (to invert accurately for the
surface). 
If a proper set of either high degree or low degree modes is not available,
the density inversion becomes very  uncertain.

Only for the MDI data were we successful in forming a local averaging
kernel that permitted a good inversion for  solar density.
Fig.~\ref{fig:diffdatagam}(a)
shows that  the difference between the density
profiles of the solar models and the helioseismologically determined
density profile is 
 $\leq 1$\% .
However, the accuracy of this measurement  is an order of
magnitude less precise than for sound speeds.

\subsection{The adiabatic index $\Gamma_1$}
\label{subsec:datagammaprofile}

The $\Gamma_1$ difference obtained between model
STD and the Sun is shown in Fig.~\ref{fig:diffdatagam}(b) .
Since most of the $\Gamma_1$ difference between the models
and the Sun is concentrated at the surface, we use only
the MDI data for the $\Gamma_1$ inversions. This set has the
most number of high degree modes. 

With the convenient  inversion method used here (SOLA), we do
not have good spatial resolution
close to the surface. There are computationally intensive inversion methods 
which give superior resolution near the surface, for example,
Optimally Localised Averages (cf. Kosovichev 1995,
Elliot \& Kosovichev 1998; Basu, D\"appen \& Nayfonov  1999).
For the general survey performed in this paper, we did not
feel it necessary to invert very close to the solar surface.

The contribution to the
solution for $\Gamma_1$ from the second function (density in this case) is
more difficult to suppress than for inversions of sound speed
or density. Thus we expect larger errors for reference models with large
density differences relative to the Sun.

Given the greater precision of the sound-speed measurements, we will
emphasize in what follows the profile of the sound speed and will only
refer to the density and $\Gamma_1$ profiles for completeness.

\section{How large are the effects of finite radial resolution?}
\label{sec:resolution}

We calculate the solar sound speed using the relative sound-speed 
difference between the models and the Sun. Thus, if
$(\delta c^2/c^2)_{\rm inv}$ is the  result of the inversion, then 
\be 
(c^2_\odot)_{\rm inv}(r_0) =
\left[{\left({\delta c^2\over c^2}\right)_{\rm inv}(r_0)}  +1 \right]
c^2_{\rm model}(r_0)  .
\label{eqn:csq} 
\ee

However, $(c^2_\odot)_{\rm inv}(r_0)$ is not  identically equal 
to the true solar sound-speed, since the 
inverted speed is an average of the sound-speed difference in the 
vicinity of 
$r_0$. The averaging kernel at $r_0, {\cal K}(r_0,r)$ defines the
region in $r$ over 
which the averaging is done. We want to estimate the 
error introduced in the inferred sound speed due to finite resolution. 
Since we do not know the true sound speed inside the 
Sun, we  estimate the errors using  
a solar model as a ``proxy Sun.'' 
Thus if $c_{\rm proxy}$ is the sound speed of the proxy Sun,  
$c_{\rm model}$ is  the  speed in the 
reference model, and $\delta c^2/c^2 = (c_{\rm proxy}^2 -c_{\rm
model}^2)/c_{\rm proxy}^2 $ the  
relative difference between the squares of
sound 	speeds of the two solar 
models , then the relative 
error in the inferred sound-speed of the proxy Sun due to the finite
resolution of the averaging kernel is 
\be 
\left({\Delta c\over c}(r_0)\right)_{\rm resol} = {1\over 2}\left[ 
\left(\int {\cal K}(r_0,r) {\delta c^2\over c^2}(r) dr\right) 
-{\delta c^2\over c^2}(r_0)
\right]. 
\label{eqn:errors} 
\ee 
The factor of $(1/2)$ in the above expression arises from the conversion 
of relative errors in $c^2$ to relative errors in $c$. 
Note that $({\Delta c/ c}(r_0))_{\rm resol}= 0 $ if the averaging
kernel $ {\cal K}(r_0,r)$ is a delta function.
The error in density due to the resolution of the density 
kernel is similarly given by 
\be 
\left({\Delta \rho\over \rho}(r_0)\right)_{\rm resol}= 
\left(\int {\cal K}_{\rm den}(r_0,r) {\delta \rho\over \rho}(r) dr\right) -
{\delta \rho\over \rho}(r_0). 
\label{eqn:errord} 
\ee 
where ${\cal K}_{\rm den}$ is the averaging kernel obtained  for density 
inversions.

In the subsequent discussions, the profile of the solar sound-speed obtained
with MDI data using STD as the reference model is referred to
as the ``standard sound speed profile''. Similarly the solar density
profile inferred  from  MDI data using STD is referred to as the
``standard density profile,'' and the solar $\Gamma_1$ profile
inferred MDI data using STD is referred to as the
``standard $\Gamma_1$ profile.''

Figure~\ref{fig:resoerr} shows $(\Delta c/c)_{\rm resol}$ for 
model NODIF and ROT when 
STD is used as the 
proxy sun.  The resolution errors are small and generally less than  
0.02--0.03 percent. 
However, the resolution errors 
are relatively large in the  solar core and the base of the 
convection zone. The large 
error in the core results from the fact that 
there are very few $p$-modes that 
sample  this region well.  This causes the averaging kernel to be
relatively wide in this innermost volume. 
The even larger resolution error at the base of the 
convection zone
is caused by the sharp gradient in the   $\delta c^2/c^2$ for the
different models. 

The rms difference between the sound speed profile inferred with the
ROT model as the reference model 
and the STD sound profile as the proxy sun 
is $0.015$\%; the rms difference  is $0.038$\% when the NODIF model is
used as the reference model. The rms difference for the density
profile is $0.080$\% when ROT is the reference model and $0.31$\% when
NODIF is used as the reference model.

The errors are larger when  NODIF is compared with STD than when  ROT
is compared with the STD model, which simply reflects 
the fact that the difference in sound speed profiles 
 between model STD and NODIF is 
larger  than the difference in sound speed profiles  between STD and ROT. 
This observation leads to the rather obvious conclusion that one 
expects to get more accurate
solar sound speeds by using reference  models which have  sound speeds
that are similar to the Sun. 
 
The errors in the inferred density are also shown in Fig.~\ref{fig:resoerr}. 
The errors in 
density are about an order of magnitude larger than the errors in the 
sound speed. 

The errors in the inferred $\Gamma_1$ due to resolution effects are 
expected to be small. 
The reason is that in the region we have been able to do the inversions
($r < 0.94$ R$_\odot$), $\Gamma_1$ differences between most models
and the Sun are very smooth (see for example Basu et al. 1999) 
The only model which could cause large
errors is model OLD.

\section{Standard sound speeds, densities and $\Gamma_1$}
\label{sec:stdsounddensity}

Table~\ref{tab:solardata} lists the  solar sound speed,  density,
and $\Gamma_1$ profiles 
that are  obtained with the STD model using MDI data. 
These data may be useful for other applications. Therefore, we have
made available 
machine readable files in the format 
of Table~\ref{tab:solardata}, but with a denser grid in radius, 
at the web-site http://www.sns.ias.edu/$\sim$jnb .

\section{Comparison with the standard solar model}
\label{sec:comparestd}

How well does the standard solar model, STD, agree with the different
measurements of the sound speed? Is the difference,
shown in Fig.~\ref{fig:diffdata}(a)-(c),  between the STD
model and each of the measurements 
larger or smaller than the
differences between the measurements 
themselves (shown in Fig.~\ref{fig:sumdiffdata})?

The root-mean-squared difference between the STD sound speeds and the
solar speeds is  0.069\% for
the MDI data, 0.069\% for the GONG data, and 0.064\% for the BiSON+LOWL
data set. 
These results are averaged over all regions of the sun for which good 
data are available, from $0.05 R_\odot$ to $0.95 R_\odot$.
In all cases, the agreement is excellent, although  the
STD model can and should be improved, especially near the base of the
convective zone (see, e.g., the discussion below
of the pre-main sequence model, PMS, and the model including rotation, ROT).
In the  solar core, where the neutrinos are produced, the rms
agreement is even slightly better: $0.062$\% for the MDI data, 
$0.061$\% for the GONG data, and $0.044$\% for the BiSON+LOWL data set. 

For all three data sets, the sound speeds   
at the base of the convection zone of the STD model 
differ by about $0.2$\% from the
helioseismological values . This discrepancy  has
been seen earlier with similar standard models
(cf. Gough et al. 1996, Bahcall et al. 1997) 
and can be attributed to the lack of mixing
in the model below the base of the convective zone. 
In the models,
diffusion without mixing causes a sudden, local rise in the
helium abundance below the base of the convective zone. 
The increase in helium abundance increases the
mean molecular weight, thereby decreasing the
sound speed (since $c^2\propto T/\mu$, where $T$ is  the local value
of the 
temperature and $\mu$ is the local mean molecular weight).

\begin{figure}
\plotfiddle{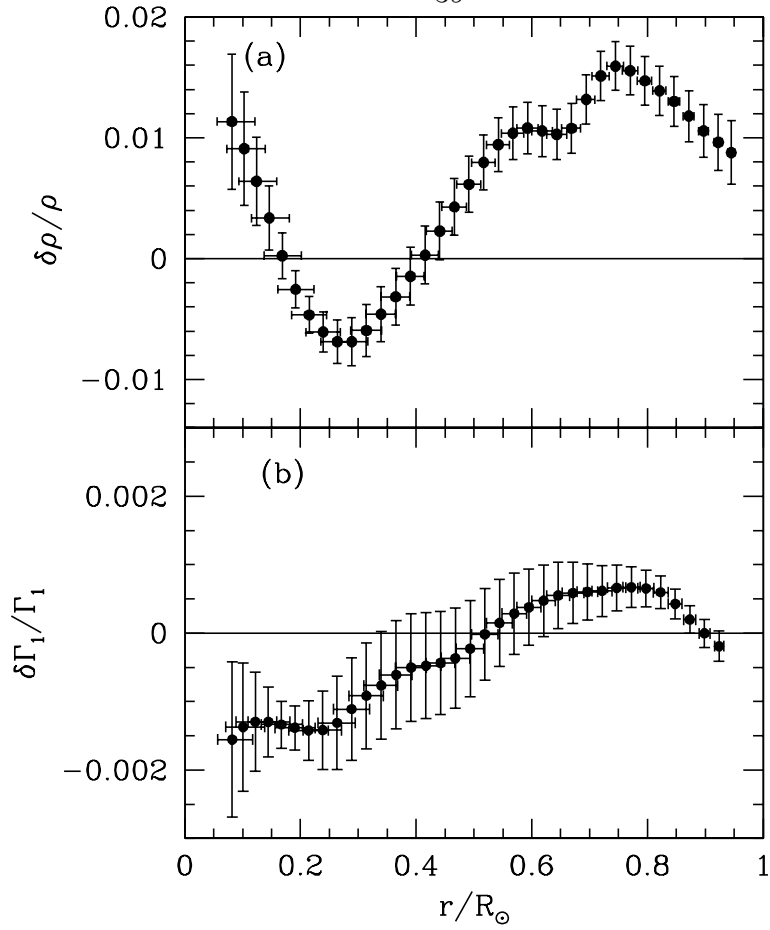}{10 true cm}{0}{90}{90}{-280}{-200}
\figcaption{ The relative density differences [Panel (a)] and
$\Gamma_1$ differences [Panel (b)] between the Sun and
model STD obtained using MDI data. The differences are in the sense 
(Sun $-$ Model)/Model. 
\label{fig:diffdatagam} } 
\end{figure}

\begin{figure} 
\plotone{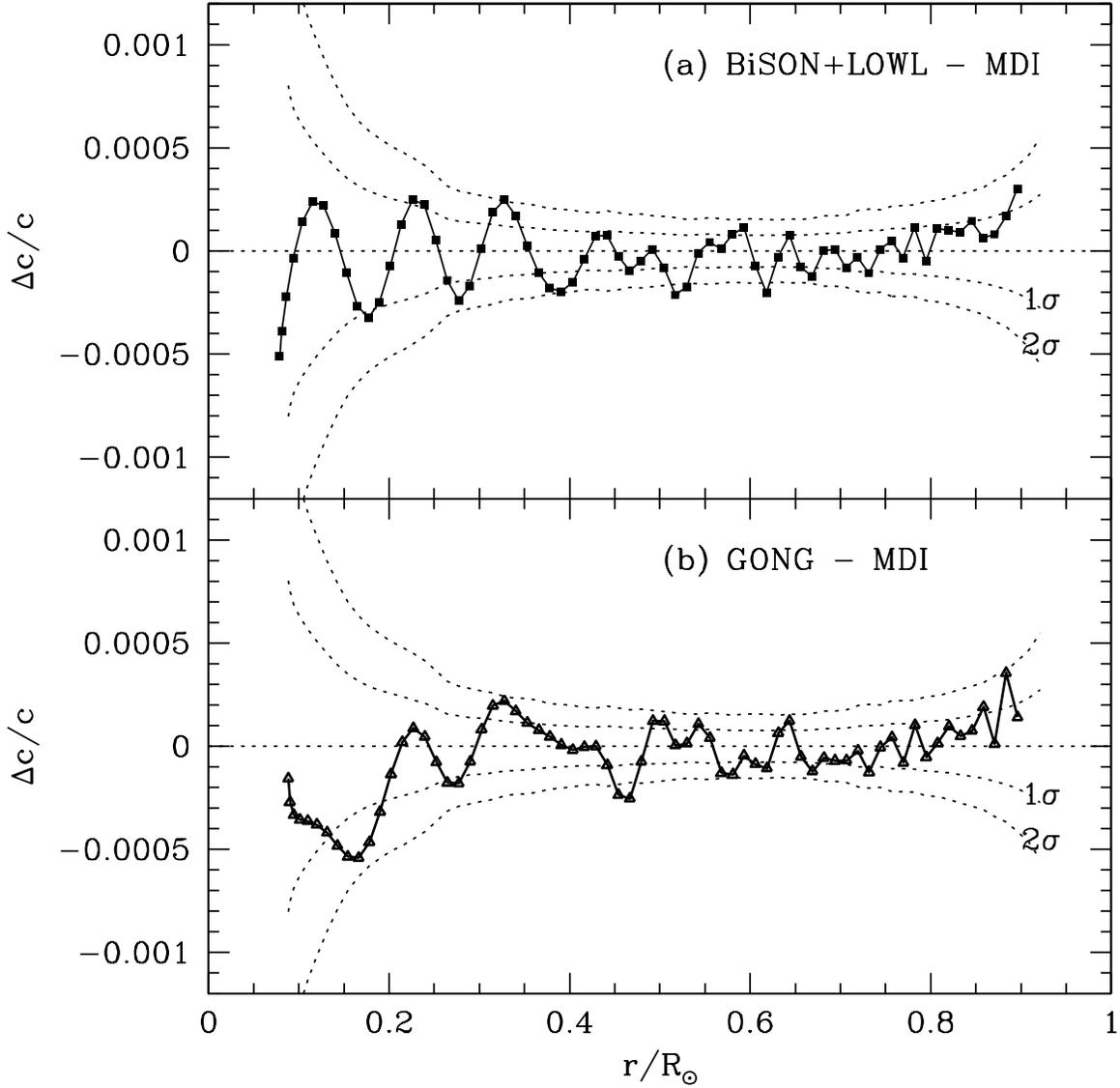} 
\figcaption{ The relative errors in the solar sound speed 
obtained with model STD as the reference model using different 
data sets. The reference sound-speed is taken to be the one 
obtained by inverting MDI data. The difference with those obtained  
using BiSON+LOWL and GONG data were calculated.  The 1 and 2$\sigma$ error 
envelopes due to errors in the data are shown as dotted lines. 
Since the reference model was the same for all three inversions, and 
since the resolution of the inversions using the three data sets  
are very similar, we do not expect any additional error due to  
finite resolution or to differences in the 
reference model. 
The rms differences are only $0.016$\% for the 
sound speeds calculated with  the BiSON + LOWL and the MDI data sets and also
$0.020$\% for the differences found between the sound speeds calculated
with the GONG and the MDI data sets.
\label{fig:sumdiffdata}
} 
\end{figure}

\begin{figure} 
\plotone{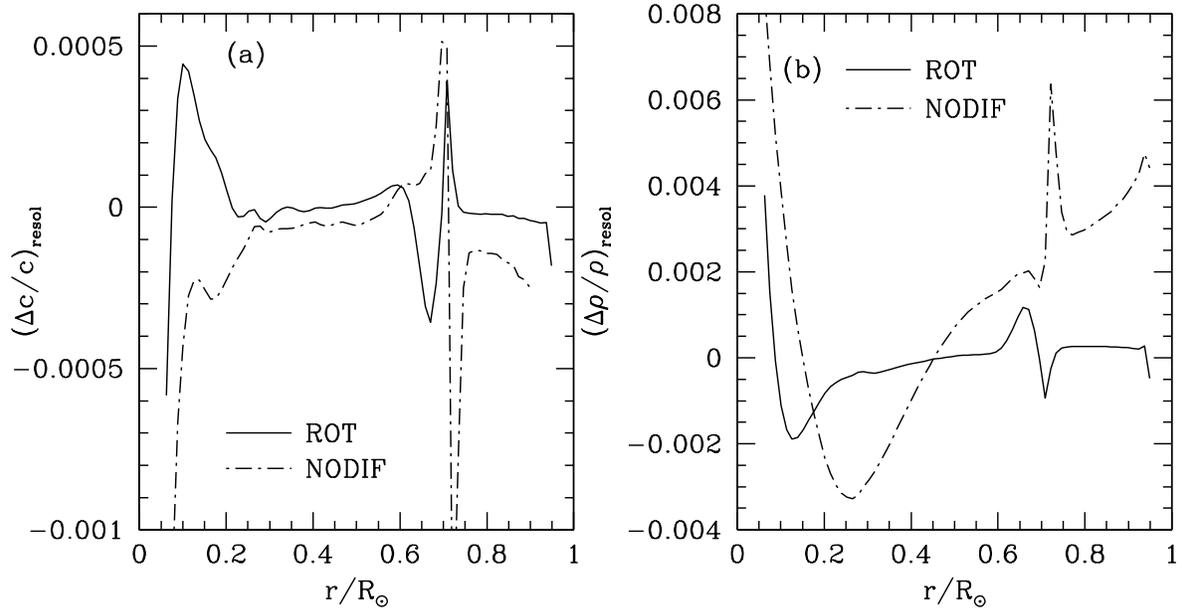} 
\figcaption{ The error in the estimated sound-speed and density of 
two models due to the finite resolution of the inversions. In both 
cases the reference models used is STD. 
The quantities $(\Delta c/c)_{\rm resol}$ and $(\Delta \rho/\rho)_{\rm resol}$,
defined in Eq.~\ref{eqn:errors} and \ref{eqn:errord}\ respectively,
are 
estimates of the errors due to resolution  that are expected in  
the solar sound-speeds 
and density obtained by inversion. The errors are largest where the 
sound-speed difference between the reference model and the test model 
shows a sharp gradient, as occurs for example, at the base of the 
convection zone. 
\label{fig:resoerr}
}
\end{figure}

The rms density difference between the model and the Sun is
within 1\% [see Fig.~\ref{fig:diffdatagam}(a)].  For
$\Gamma_1$, the rms difference between the STD model and the
Sun is less than 0.1\%  [Fig.~\ref{fig:diffdatagam}(b)].
 The somewhat larger  difference in the
core has been attributed to errors in the equation of state
(cf., Elliott \& Kosovichev 1998).

\section{Comparison with variant and deficient models}
\label{sec:compvariants}

How well do the variant models discussed in \S~\ref{subsec:variants} 
and \S~\ref{subsec:deficient} agree with
the helioseismological measurements?

Figures~\ref{fig:cdiffmdi} and \ref{fig:cdiffbest} show the results of
the inversions made using, respectively, MDI (Fig.~\ref{fig:cdiffmdi})
and GONG and BiSON+LOWL (Fig.~\ref{fig:cdiffbest})
data. For most of the models, the vertical scale for the fractional
velocity differences has a range of a few tenths of a percent. For the
densities, the corresponding range is of order a few percent.

For $\Gamma_1$, Fig.~\ref{fig:gamdif} shows the fractional differences
between the Sun and the solar models that were found using MDI data.
The fractional differences are less than or of order of $0.2$\%
for all of the solar models except for the MIX model. For the MIX
model, the fractional differences are larger, of order $0.5$\% .  We
conclude that, with the exception of the MIX model,  the 
theoretical profiles of $\Gamma_1$ are in good agreement with the
solar values of $\Gamma_1$.

Table~\ref{tab:rms} summarizes  the average 
root-mean-squared deviation between
the predicted sound speed profile of the 
 solar models discussed in \S~\ref{sec:models} 
and the measured sound profile determined with the MDI data.

\subsection{Variant models}
\label{subsec:comparevariant}

\begin{figure} 
\plotone{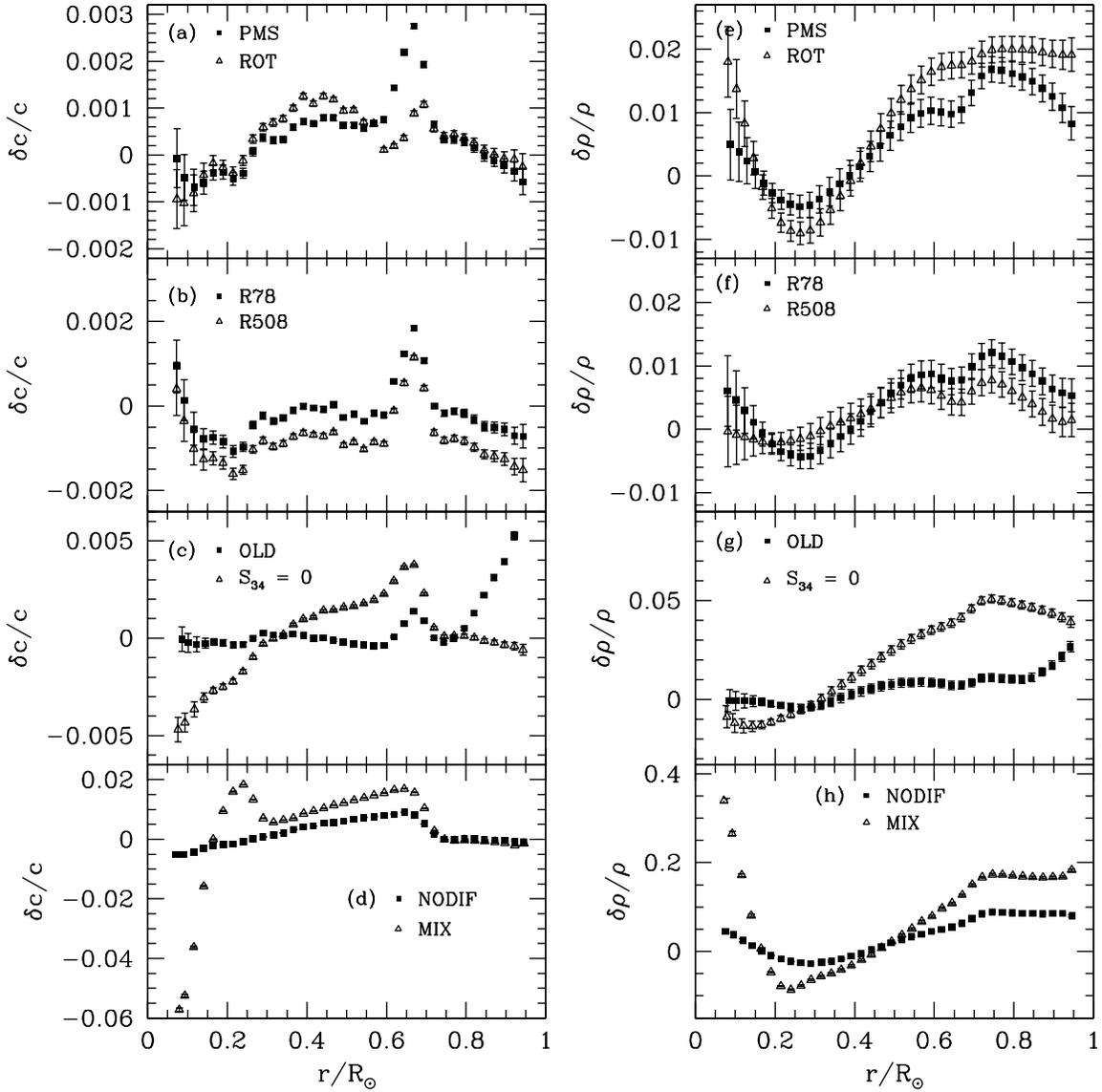} 
\figcaption{The relative sound-speed and density differences 
between the Sun and different solar models obtained using MDI data. 
The differences are in the sense (Sun $-$ Model)/Model. 
Horizontal error-bars are not shown for the sake of clarity. 
The differences are larger for the models described in the bottom two
sets of panels and therefore the vertical scales cover a wider range
for these panels.
The different models are described in Table~1. 
\label{fig:cdiffmdi}
}
\end{figure}

\begin{figure} 
\plotone{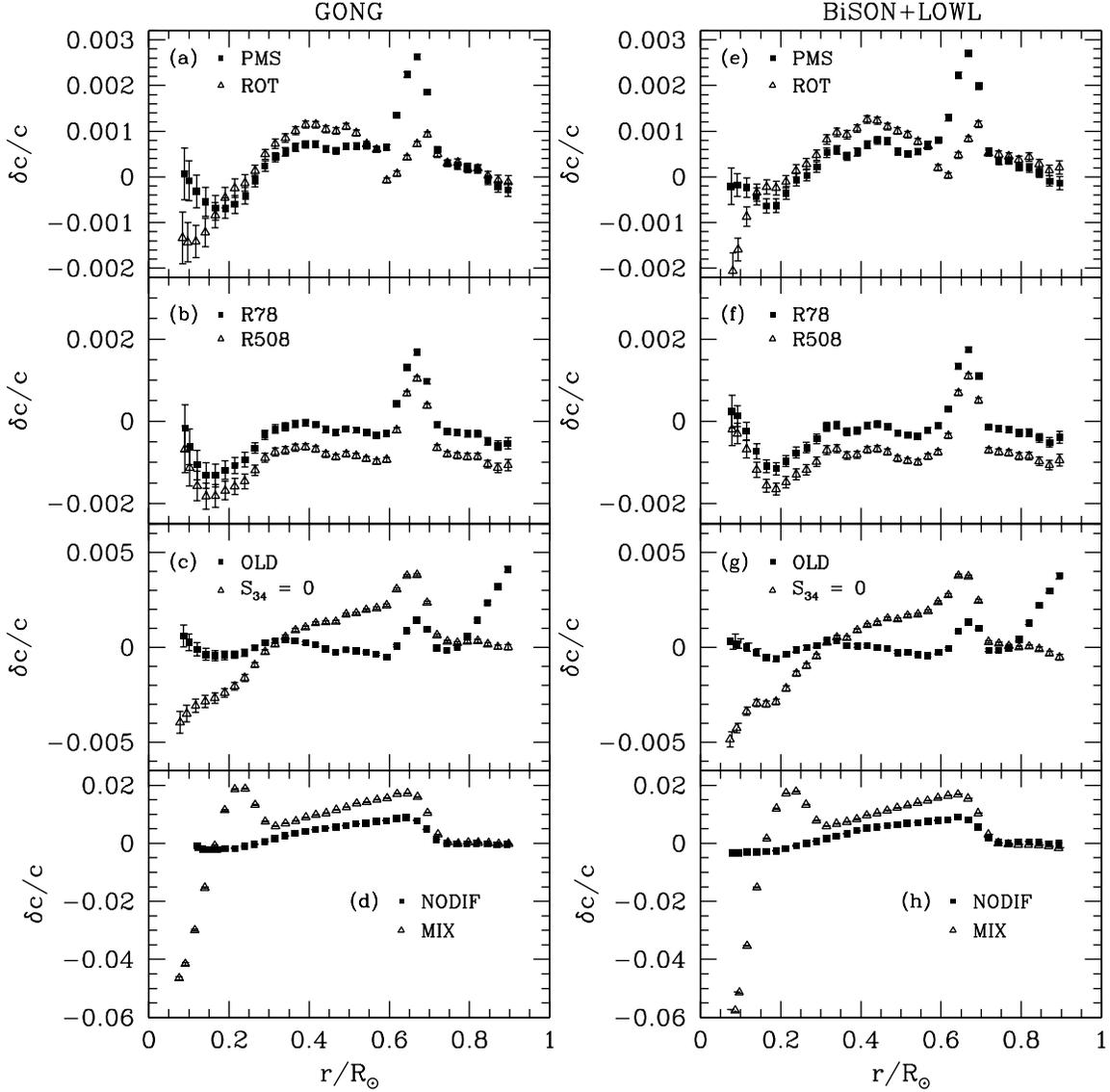} 
\figcaption{The relative sound-speed differences
between the Sun and different solar models obtained using
GONG and BiSON+LOWL data.
The differences are larger for the models described in the bottom two
sets of panels and therefore the vertical scales cover a wider range
for these panels. The different models are described in Table~1.
\label{fig:cdiffbest}
} 
\end{figure}

\begin{figure} 
\plotone{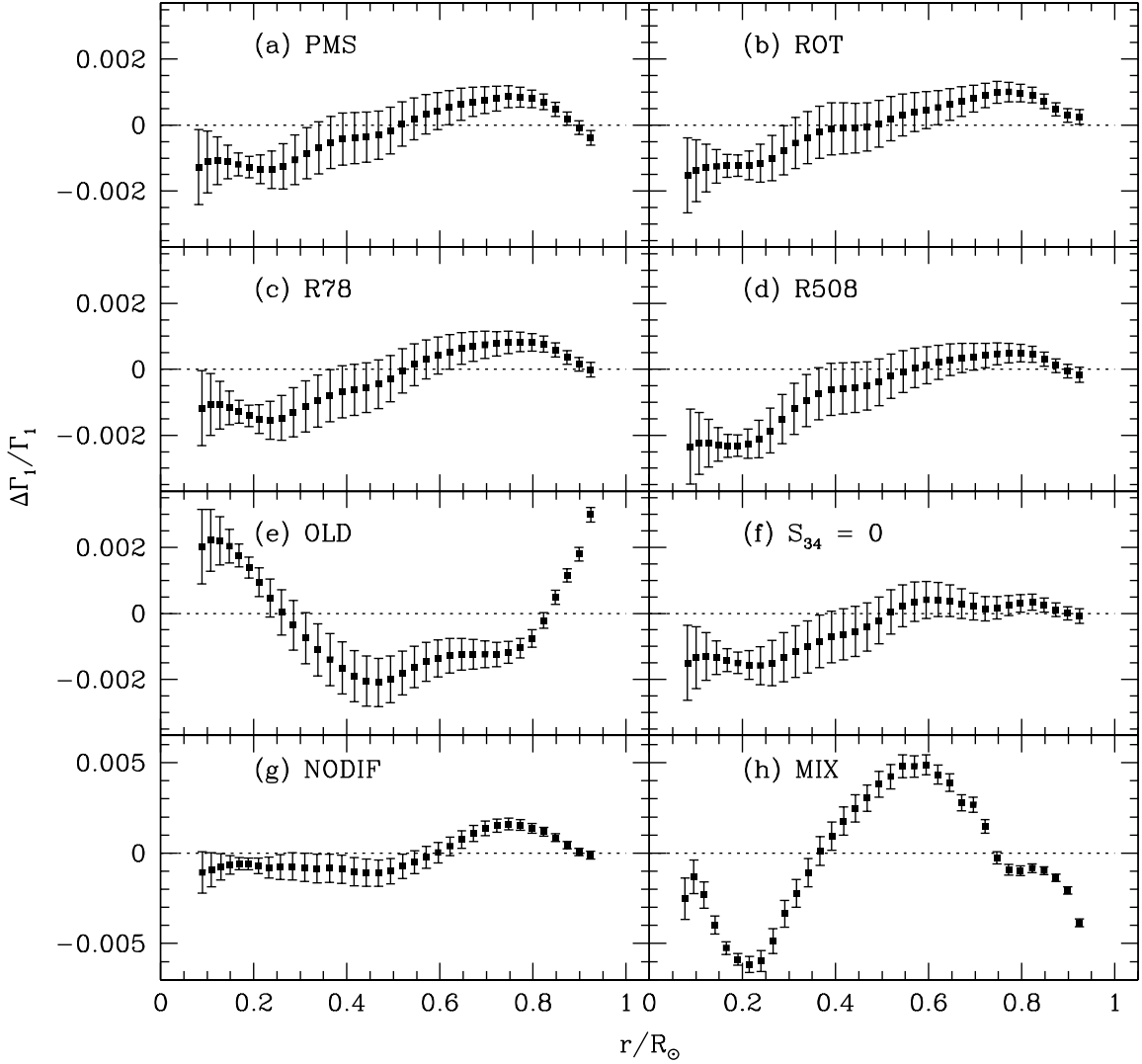} 
\figcaption{The relative $\Gamma_1$ differences
between the Sun and different solar models obtained using
MDI data.  This figure is similar to Fig,~\ref{fig:cdiffmdi} except that
Fig.~\ref{fig:gamdif} refers to $\Gamma_1$ instead of sound-speed and density.
The vertical scale of the last two panels of  
Fig.~\ref{fig:gamdif} is larger than the other panels in order 
to accommodate the
large difference for model MIX [Panel (h)]
\label{fig:gamdif}
} 
\end{figure}

The pre-main sequence PMS model yields values for $\delta c/c$ that
are  similar to the results obtained with the STD model
(comparing Fig.~\ref{fig:cdiffmdi}a and Fig.~\ref{fig:cdiffbest}a,e with 
Fig.~\ref{fig:diffdata}a-c).  This similarity is to be expected
since the difference between
the STD and PMS models is  small, of order 
hundredths of a percent in $\delta c/c$ everywhere and of order a few
tenths of a percent in $\delta \rho/\rho$
(cf. Fig.~\ref{fig:diffmodels}a,d). 
The difference in $\Gamma_1$ between the
PMS model and the Sun is similar to that between the STD and the
Sun [cf. Fig.~\ref{fig:diffdatagam}(b) and Fig.~\ref{fig:gamdif}(a)].

The model with rotational mixing, ROT, agrees better than the STD
and PMS models with the helioseismological measurements near the base of the
convective zone (see Figures~\ref{fig:cdiffmdi}a,e and
\ref{fig:cdiffbest}a,e). This improved agreement
confirms the suggestion (see Richard et al. 1996) 
that mixing at the base of the
convection zone is a possible explanation for the significant
discrepancy in this region between the measured and the STD model
sound speeds. However, with this version of mixing, the agreement is
slightly worse in the solar core, resulting in an overall rms
deviation that is essentially the same as for the STD model.
The difference in  $\Gamma_1$ between ROT and the Sun is very similar to
differences in $\Gamma_1$ found with 
models STD and PMS.

The two models with slightly different radii, R78 and R508, yield
results (see Table~\ref{tab:rms})
for the rms agreement with the MDI data that are comparable to
the STD model.  The R78 model yields slightly better agreement and the
R508 model yields slightly worse agreement than is obtained with the
standard model radius. 
The shape of the sound speed 
differences between the R78 and R508  models and the Sun is very
similar to the shape of sound speed differences 
between the  STD model and the Sun, but for the `non-standard' radii
the sound speeds are shifted downward in Fig.~\ref{fig:cdiffmdi}b
and Fig.~\ref{fig:cdiffbest}b,f (see also Fig.~\ref{fig:cerrora}c,d).

\subsection{Comparison with deficient models}
\label{subsec:comparedeficient}

The model constructed using old input data, OLD, produces a
significantly worse rms sound speed discrepancy, $0.17$\%, compared to
the $0.07$\% for the STD model(see Table~\ref{tab:rms}).  It is
encouraging that the improvements in nuclear physics, equation of
state, and radiative opacity that are described in \S~\ref{subsec:std}
have resulted in better agreement, by about a factor of two, 
 with the measured sound speeds.
The OLD model also shows  larger difference, relative to the Sun,  in $\Gamma_1$ 
towards the surface, confirming our suspicion that the equation of state used
 in the OLD model is not sufficiently accurate for optimal 
helioseismological applications.

The model with S$_{34} = 0$ does not allow the nuclear reaction 
$^3$He($\alpha$,$\gamma$)$^7$Be . This change in the nuclear physics
results in a sufficiently large modification  in the structure of the solar
core in the model that the difference is easily seen in precise
helioseismological measurements (see Bahcall et al. 1997). 
 Figures~\ref{fig:cdiffmdi}c and
\ref{fig:cdiffbest}c,g show that the sound speed predicted by  the 
S$_{34} = 0$ model differs from the helioseismologically inferred
sound speeds by as much as $0.5$\% in the solar core, about an order
of magnitude worse agreement than is obtained for the core with the STD model.
The influence of the reaction is seen even more dramatically
in the density differences with respect to the Sun. 
Figure~\ref{fig:cdiffmdi}(g) shows that the discrepancies in density are
as large as $5$\% in the outer region of the sun. This result can be
understood as follows.  To achieve the same luminosity, the density in
the core must be increased when an important nuclear reaction,
$^3$He($\alpha$,$\gamma$)$^7$Be, is artificially set equal to zero.
Since 
mass is conserved, any change in the density in the core has to be
compensated by an opposite, larger  change in the less dense outer layers.
The  small reduction  relative to the STD model in the density in the 
core of the S$_{34} = 0$ solar model  
results in a
relatively large change in the density in the outer layers and a
significant discrepancy with the helioseismologically inferred density
profile.
Since the equation of state used in this model is the same as
in STD, the differences in $\Gamma_1$
between this model and the Sun are very similar to those
found with the STD model.

The model NODIF is a generally poor fit to the helioseismological
measurements. Figures~\ref{fig:cdiffmdi}d,h  and
\ref{fig:cdiffbest}d,h show that the disagreement is consistently
large near the base of the convective zone, which reflects the fact
that omitting diffusion results in models with  incorrect depth
of the convection zone (Bahcall \& Pinsonneault 1995,
cf. also Table~\ref{tab:diffmodels} of the present paper). 
The rms discrepancy between the sound speeds of
the NODIF model and the measured sound speeds is $6.5$ times worse on
averaged over the sun than for the STD model. 
 In the region where we have inverted for $\Gamma_1$ difference,
NODIF fares quite well. The main difference in $\Gamma_1$ between the NODIF
model
and the Sun is expected to arise because of differences in 
helium abundance. But that difference will show up   only in the helium ionization
zone (around $ 0.98R_\odot$) which we have not resolved.

The vertical scale for $\delta c/c$ must be increased by a factor of
$25$, from $0.002$ to $0.05$(negative discrepancy), 
in order to display the very large
discrepancy that exists for the MIX models.
The differences are particularly 
glaring in the
core, where the model is fully mixed. The sound speed difference
between the MIX model and the measurements 
is as much as 5\% in the solar core 
while the density difference is almost 40\%. 
The corresponding maximal differences for the standard model in the
solar core are $0.1$\% in $\delta c/c$ and $1$\% in $\delta
\rho/\rho$, about $50$ times and $40$ times smaller than for the MIX
model. The average rms discrepancy for the MIX model is about $25$ times
larger than for the STD model. Obviously, this model is not a good
model of the sun, although it was proposed (Cumming \& Haxton 1996) 
as a way of decreasing (but
not eliminating) the differences between standard neutrino flux
predictions and the measured neutrino fluxes.
We were unable to obtain a reliable inversion for $\Gamma_1$
with model MIX because of the large difference in density between
the model and the Sun.

\section{Dependence upon the assumed reference model}
\label{sec:dependence}

In this section, we evaluate the dependence of the inferred sound
speed profile and the inferred density profile upon the assumed
reference model. 
We calculate the sound speed or the density using two
different combinations of reference model and measurement 
data and then compare
the results.  Thus we evaluate the set of differences formed by 
(model$_i$ - data$_k$) - (model$_j$ - data$_j$), where for convenience we
always take model$_j$ as the STD model and data$_j$ as the MDI data
set, but model$_i$ can be any one of the eight variant models
discussed in \S~\ref{subsec:variants} and data$_k$ is either the
MDI, BiSON+LOWL, or GONG data set.

The principal results of this section are summarized in
Table~\ref{tab:tabdif}.

\subsection{Dependence of sound speed profile upon reference model}
\label{subsec:speeddependence}

\begin{figure} 
\plotone{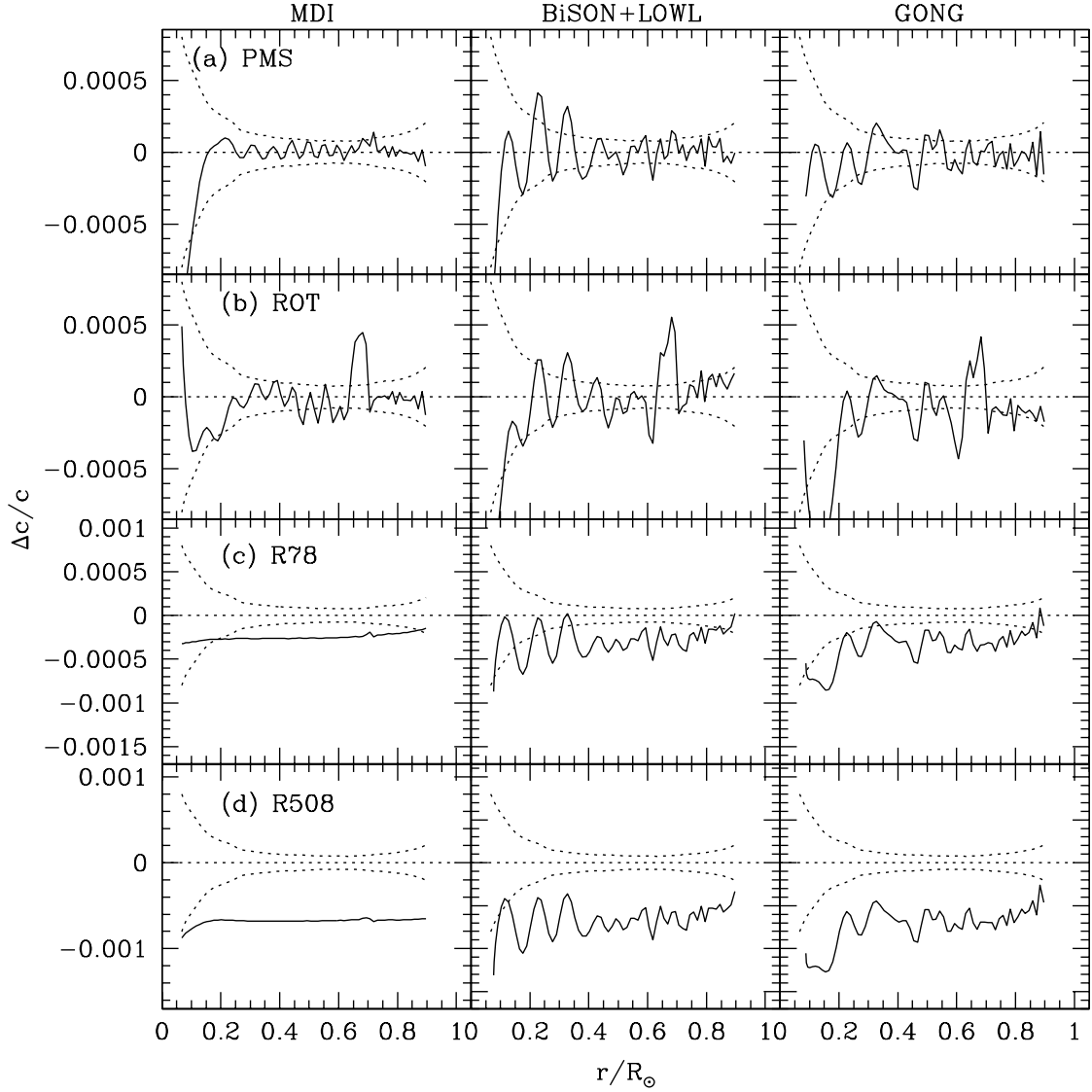} 
\figcaption{The fractional difference between  the solar sound speed 
inferred  using  the STD solar model as a reference model and 
the sound speed obtained using each of four variant solar models
as a reference model. For the reference sound profile, the MDI data
set was inverted.
Differences in the inferred sound speeds are presented for all three
data sets, MDI, BiSON+LOWL, and GONG, and for all of the variant solar
models. The $1\sigma$  
error envelope due to data measurement  errors is shown as the dotted line. 
\label{fig:cerrora}
} 
\end{figure}

\begin{figure} 
\plotone{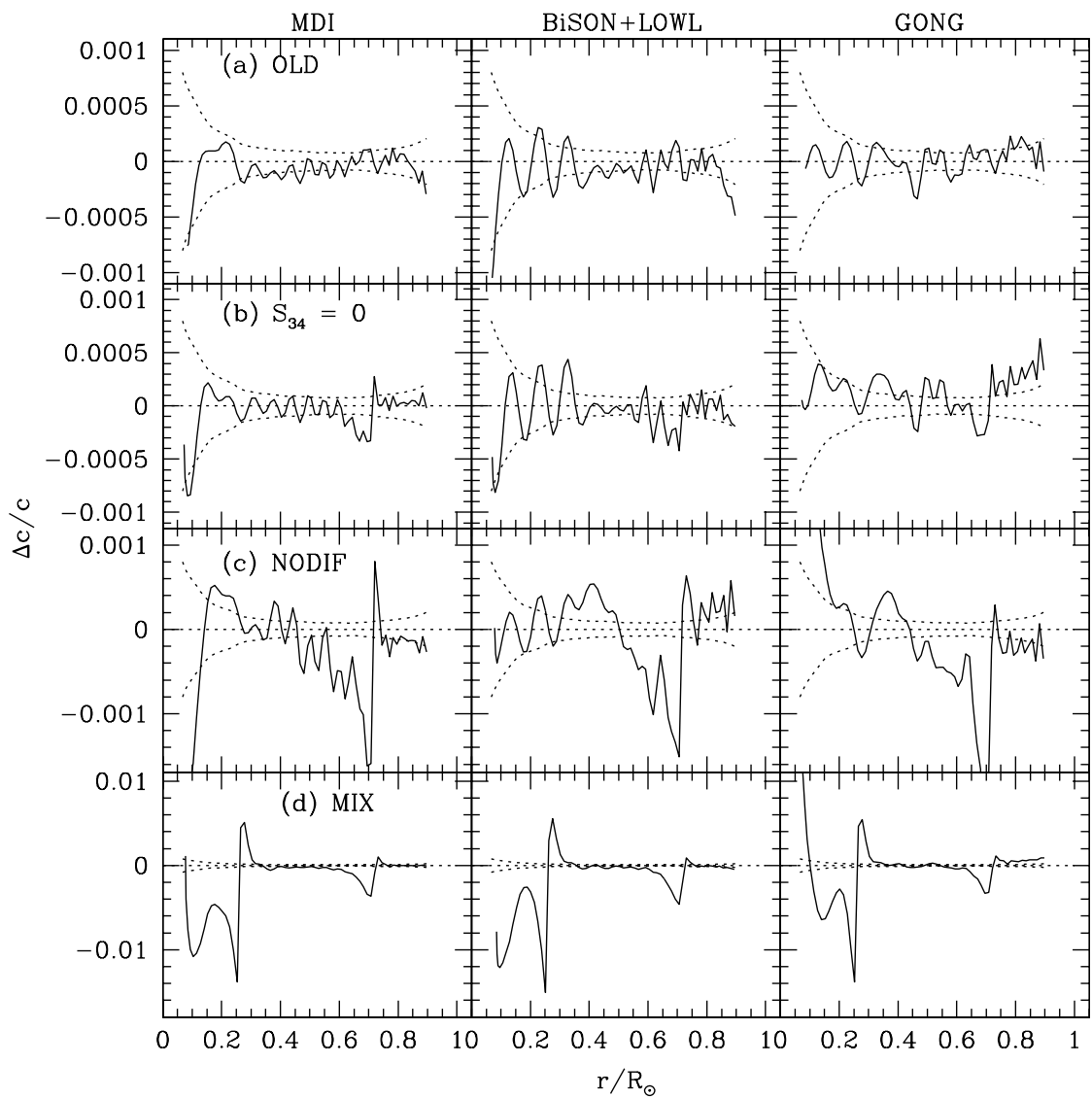} 
\figcaption{ The same as Fig.~\ref{fig:cerrora}, but for four solar models
with considerably different physics than the STD model. 
The fractional differences obtained using the MIX model as the
standard model are an order of magnitude larger than for the other
cases considered; therefore, the vertical scale for the bottom panel
covers an order of magnitude larger range. 
\label{fig:cerrorb}
} 
\end{figure}

Figures~\ref{fig:cerrora} and \ref{fig:cerrorb} show the relative
differences (double differences in the sense defined above) 
between the  standard sound speed profile (obtained with the model STD and MDI 
data ) and the
solar sound speed inferred  using the eight variant models.
We present result for all three of the data sets: MDI, BiSON+LOWL, and
GONG. 

The fractional difference, $\Delta c/c$,  
that is shown in Fig.~\ref{fig:cerrora} and Fig.~\ref{fig:cerrorb} is
defined explicitly by the relation
\begin{equation}
{\Delta c \over c} = {{c_{\odot,~variant} - c_{\odot,~STD}} \over
c_{\odot,~STD}} .
\label{eqn:defnDeltac}
\end{equation}
The quantities $c_{\odot,~variant}$ and $c_{\odot,~STD}$ are the best
estimates for the solar sound speed that are found 
 using the specified reference model
and data set.  More explicitly,
$c_{\odot,~model} = c_{\rm model}\sqrt{(1 + \delta c^2/c^2)_{\rm model}}$,
 where $(\delta c^2/c^2)_{\rm model}$ is obtained directly
from the inversion equation, Eq.~(\ref{eqn:inv}).

The inversions  obtained using the PMS, the ROT, the OLD, and the 
S$_{34} = 0$
models all yield sound
speeds that differ from the standard sound speed profile by only a few
hundredths of a percent, i. e., within the errors due to the measurements
(see Table~\ref{tab:tabdif} and Fig.~\ref{fig:cerrora} and
Fig.~\ref{fig:cerrorb}). The
only conspicuous exception to this statement occurs near the base of the
convective zone, where the finite resolution causes a difference that
is larger than the recognized measurement errors (see
Fig.~\ref{fig:cerrora}b).

The largest  systematic,  monotonic differences, $\sim 0.03$\% to
$0.08$\%, are
found for the models 
R78 and R508(see Fig.~\ref{fig:cerrora}c,d), although other models
have larger rms differences(cf. Table~\ref{tab:tabdif}). 
The  sound-speed profiles obtained
 using these  two models as reference models (and the MDI data for the
measurements) 
exhibit a
smooth difference with respect to the standard sound-speed profile.
Data errors and resolution errors
are not important when the MDI data are used for both the standard and
the variant model inversions.  One can see the irregular effects of
using different data sets with finite resolution in the BiSON+LOW and
GONG panels of Fig.~\ref{fig:cerrora}c,d.
Equation~(\ref{eqn:freqdif}) is obtained assuming that there
is no difference in radius between the Sun and the model; hence, 
a difference in the radius between the model and the sun 
can introduce a systematic error in the inversion results.
One must use an accurate value for the solar radius in order to obtain
precisely correct inversion results (see
Antia 1998 and Basu 1998).

Even when the NODIF model is used as the reference model, the inferred
sound speed profile is in reasonable agreement with the standard
profile except near the base of the convective zone. At the base of
the convective zone,  the
combined effect of finite resolution and non-linear effects increase
the velocity difference 
well beyond what is expected from measurement errors alone. 
The results for the NODIF model are shown in Fig.~\ref{fig:cerrorb}c.

The linear inversion fails for the MIX model in the solar core and
near the base of the convective zone (see Fig.~\ref{fig:cerrorb}d). 
The rms fractional differences, $\Delta c/c$, are 
$\sim 0.4$\% (see Table~\ref{tab:tabdif}), 
an order of 
magnitude larger for the MIX model than for the models which more
closely resemble the STD model.
This failure is not surprising since the MIX model is very different
from the sun (and the STD model) in the core and at the base of the
convective zone.
What is more remarkable is that despite the very large 
difference between the reference model and the Sun, the
linear inversion scheme yields sound speeds for much of the solar volume 
that are within a 
few tenths of a percent of the results obtained using much better 
reference models.

\subsection{Dependence of the inferred density profile upon reference model}
\label{subsec:densitydependence}

\begin{figure} 
\plotone{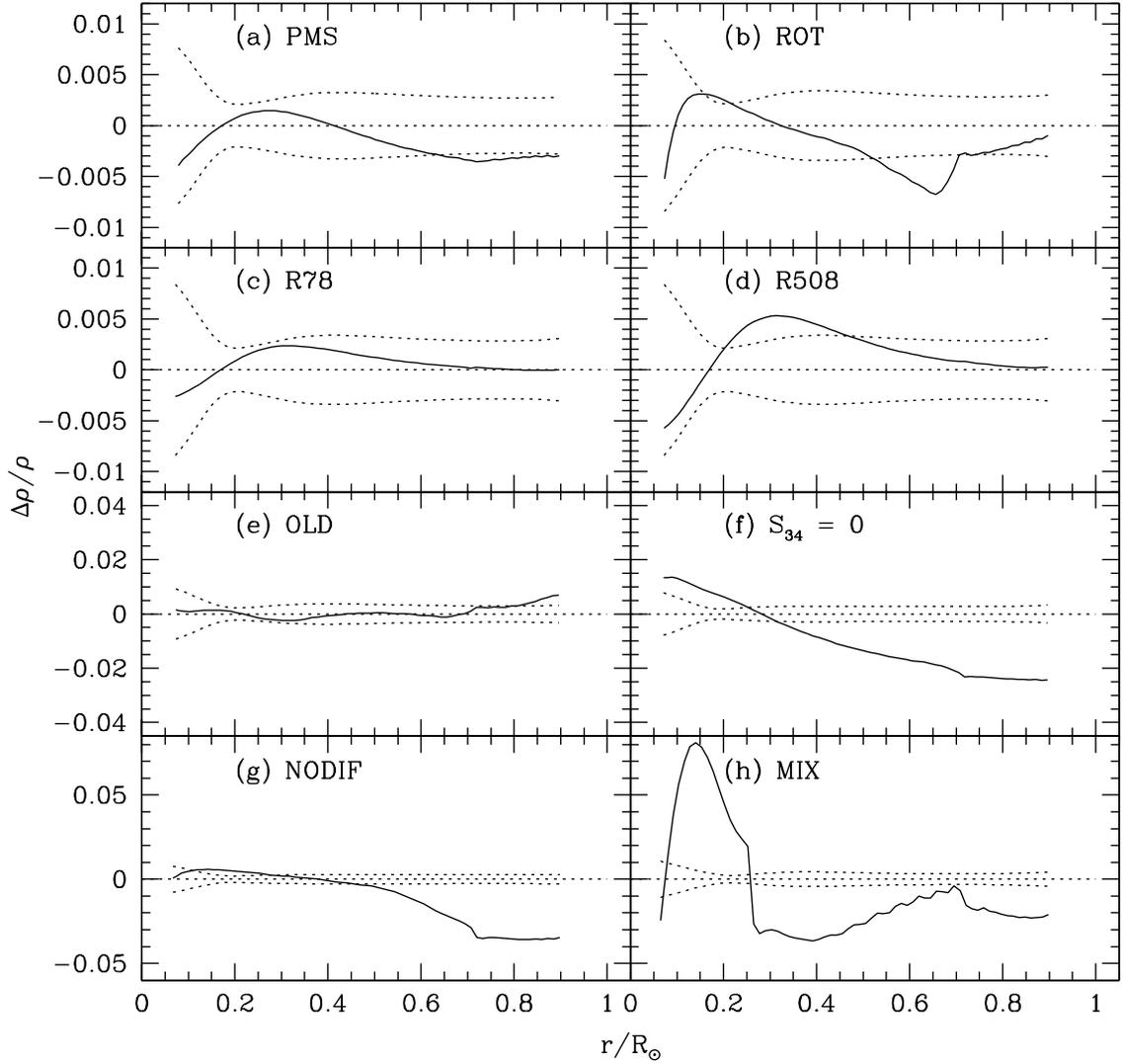} 
\figcaption{The relative errors in the solar density 
obtained using different reference models and the MDI 
data set. The reference solar density used to 
compute the differences is the one obtained  
using model STD as the reference model. The $1\sigma$  
error envelope due to data errors is shown as the dotted line. 
The differences are larger for the models described in the bottom two
sets of panels and therefore the vertical scales cover a wider range
for these panels.
\label{fig:derror}
} 
\end{figure}

Figure~\ref{fig:derror}  shows the dependence of the inferred density
profile upon the assumed reference model. We are able to make these
comparisons only for one data set, the MDI data set, since we were not
able to make  satisfactory inversions for the density profile using the
other data sets.

For the PMS, ROT, R78, R508, and OLD models, which differ from each
other only by modest amounts, the dependence of the inferred density
profile upon the assumed reference model
is moderately large, of order $0.2$\% (see the next-to-the-last column of 
Table~\ref{tab:tabdif}), 
but  is nevertheless 
generally smaller than  the estimated
measurement uncertainties.
Of course, this dependence upon reference model 
is about an order of magnitude larger than
the dependence of the sound speeds upon reference models (see above).

For the NODIF, S$_{34} = 0$, and MIX models, the dependences shown in
Fig.~\ref{fig:derror} are much larger than the measurement errors, 
i.e. they are $\sim$ a few percent. 
Non-linear effects are clearly important in
these inversions. Nevertheless, the most remarkable fact may be that,
despite the very significant differences between the variant models (NODIF,
S$_{34} = 0$, and MIX) and the STD model, the different models all
yield estimates for the solar density that agree with each other within
a few percent.

\subsection{Dependence of the inferred $\Gamma_1$ profile upon reference model}
\label{subsec:gammadependence}

\begin{figure} 
\plotone{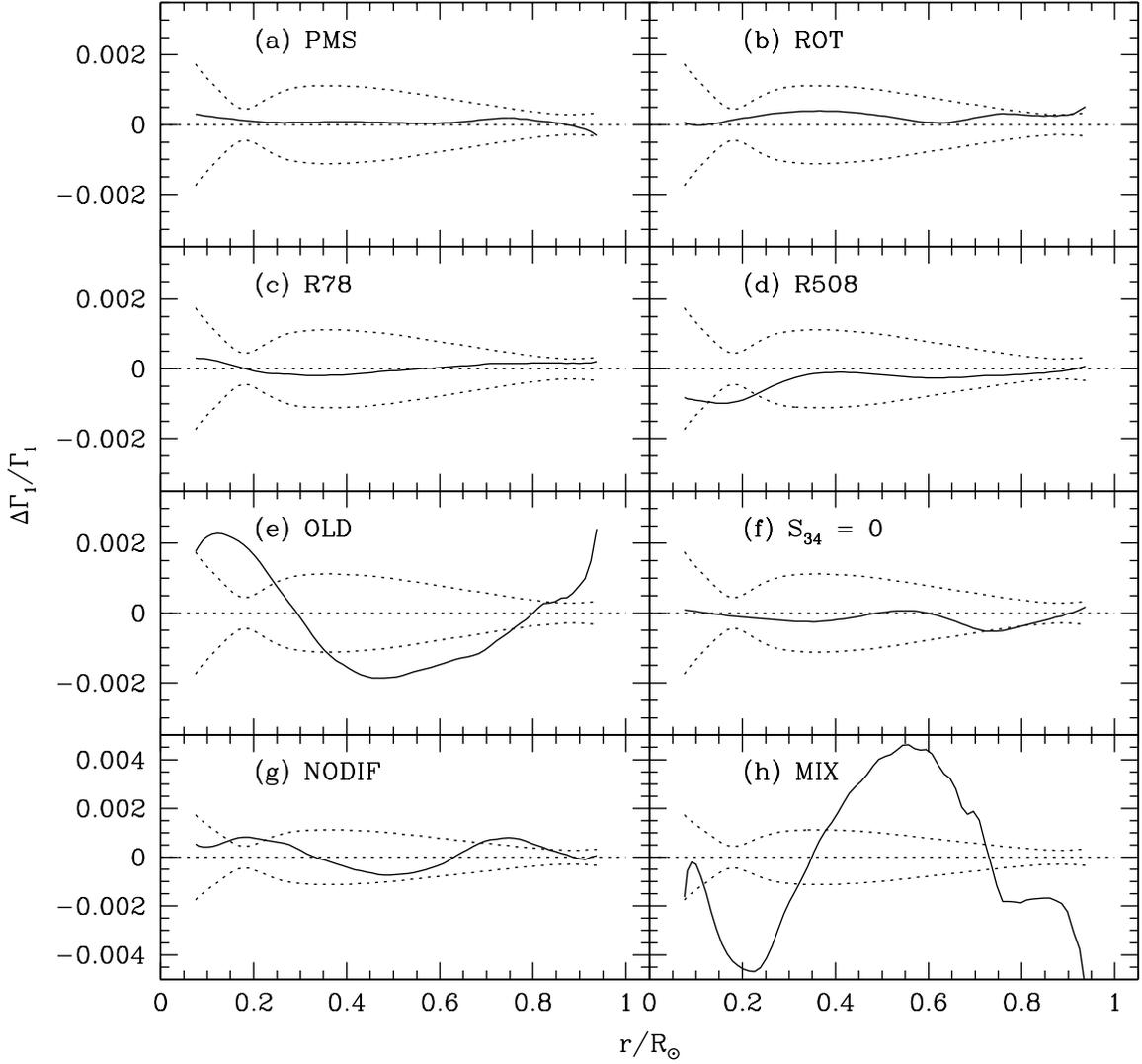} 
\figcaption{The relative errors in the adiabatic
index $\Gamma_1$ for the Sun.
obtained using different reference models and the MDI 
data set. The reference solar density used to 
compute the differences is the one obtained  
using model STD as the reference model. The $1\sigma$  
error envelope due to data errors is shown as the dotted line. 
The differences are larger for the model described in the last
panel (h)
 and therefore the vertical scales cover a wider range
for  panels (g) and (h). 
\label{fig:gerror}
} 
\end{figure}

Figure~\ref{fig:gerror}  shows the dependence of the inferred $\Gamma_1$
profile upon the assumed reference model. All the comparisons are
for the MDI data set.

For models PMS, ROT, R78, R508 and S$_{34} = 0$, the $\Gamma_1$ 
profiles obtained
agree well with the Sun and with each other 
( $\sim$ of a few hundredths of a percent, see the last column of 
Table~4 for the rms 
dependence upon the reference model). 
Although the
structure of the S$_{34} = 0$ model is quite different from the
Sun (and model STD), the structural difference is not
large enough to cause major problems with the
$\Gamma_1$ inversion.
The $\Gamma_1$ profile obtained  with
model NODIF shows a moderately large difference, 0.05\%.
 The profile obtained with model OLD shows an even larger difference,
which we believe is due to an inadequate description of the
structure close to the solar surface. 
By far the largest differences are found for model MIX, which
is due to the large
difference in structure between this model and the Sun.

\section{Discussion} 
\label{sec:discussion}

The principal purpose of this paper is to explore some of the systematic
uncertainties that affect the determination of the profiles of the
solar sound speed and the solar density.

As a by-product of this investigation, 
we have confirmed that standard solar models are in 
remarkable agreement with  helioseismological measurements of the
sun. 
For example, 
the rms difference  between the standard solar model profile for
sound speeds and the helioseismological profile is only $0.07$\% (see
discussion in \S~\ref{sec:comparestd}). Including pre-main sequence
evolution or a small amount of rotationally induced mixing does not
affect the average results very much, but 
can give better agreement with observations near the base of the
convection zone(see
discussion in \S~\ref{sec:compvariants}.

Table~\ref{tab:rms} shows that 
five state-of-the-art solar models(STD,PMS,ROT,R78,R508), each constructed 
with some different physics
or parameter choice, all give comparable agreement with the
global helioseismological measurements.
On the basis of the global seismological evidence, one cannot say that
one of these models is definitely more like the sun than the other models.

\subsection{Systematic uncertainties}
\label{subsec:systematics}

\subsubsection{Uncertainties in the data}
\label{subsubsec:datauncertainties}
We determine in \S~\ref{sec:datadiffs}
the systematic differences due to the choice of the
individual data set by comparing sound speed profiles calculated using
different data sets.  The results are shown in Fig.~\ref{fig:sumdiffdata}; the
difference between the results from state-of-the-art data sets is rms
about $0.02$\% averaged over the sun 
and bounces around within the $2\sigma$ error envelopes
determined by the combined measurement errors.

\subsubsection{Effects of finite resolution}
\label{subsubsec:finteresolution}
We estimate in \S~\ref{sec:resolution} the uncertainties due to 
the finite resolution of the inversion kernel 
by adopting a particular solar model as a proxy sun and then comparing
the convolved and inverted sound speed (or density) profile with the
true profile in the proxy sun.  The finite resolution of the inversion
kernel
leads to rms systematic uncertainties in the range 
$\sim 0.02$\% to $0.04$\% in the profile of the sound speed, 
although the errors are typically much larger in the solar core
and at the base of the convective zone( see Fig.~\ref{fig:resoerr}).
The uncertainties in the density profile due to finite resolution are
typically an order of magnitude larger than the errors in the profile
of the sound speed.

\subsubsection{Uncertainties due to reference models}
\label{subsubsec:referenceuncertainties}
We use nine different solar models in order to determine the effects
of the choice of reference model upon the inferred sound speed, 
density and $\Gamma_1$ profiles (see the discussion in sect~\ref{sec:dependence}). 
The results are summarized in
Table~\ref{tab:tabdif}.

We  have performed calculations for  a standard solar 
model (STD) and four variant
models(PMS,ROT,R78, and R508, each described in
\S~\ref{subsec:deficient}). 
All five of these models include
physics and input parameters that are at the state-of-the-art for
$1999$ solar models. The average rms difference between the sound velocities
of each of the variant models and the STD model is $0.07$\% (see
\S~\ref{subsec:variants}). The average rms difference between the sound
speed profile inferred for the Sun using one of the variant models and
the STD model is $0.03$\%(averaging the first four lines of 
Table~\ref{tab:tabdif}).  
Hence,
the spread among the inferred solar sound
speeds is more than a factor of two less than the spread among the
reference models themselves.

We  also performed calculations for  four deficient 
models(OLD, $S_{34} = 0$, NODIF, and MIX). The physics used in
constructing each of these models is deficient in some significant way
(see \S~\ref{subsec:deficient}).
These deficiencies are reflected in the fact that the 
average rms difference between the sound velocities
of each of the four non-standard models and the STD model is $0.7$\% (cf.
\S~\ref{subsec:deficient}), an order of magnitude larger than for the
variant models.  Nevertheless, these deficient-by-design models give,
when used as reference models, reasonably accurate values for the
inferred solar sound speeds. The 
average rms difference between the sound
speed profile inferred for the Sun using one of the deficient  models and
the STD model is $0.13$\%(averaging the last four rows of 
 Table~\ref{tab:tabdif}).  Thus
the discrepancy, when averaged over the different deficient models, is
a factor of more than five less than the spread among the
reference models.

Our bottom line on the systematic uncertainties for sound speeds 
is that, as expected, even
relatively crude reference models yield reasonably good estimates for
the solar sound speed.  

Table~\ref{tab:rms} and Table~\ref{tab:tabdif}
show that the profile of $\Gamma_1$ is determined with a
precision that is similar to, or slightly better than, the profile of
the solar sound speed, that is to an accuracy  $\sim 0.1$\%.  
The density is determined with an order of
magnitude less precision, $\sim 1$\%.

\subsection{Can helioseismology rule out some non-standard solar models?}
\label{subsec:ruleout}

What models are strongly disfavored (ruled out) by the
helioseismological data? We choose as a figure of merit (crudely analagous
perhaps to $1$ standard deviation) the largest  rms difference
in sound speed profile between the Sun and one of the variant
(state-of-the-art) solar models. This rms difference is $0.1$\% (for
the R508 model, see Table~\ref{tab:rms}). The $S_{34} = 0$ model has approximately twice as
large a deviation($0.2$\%) and is therefore somewhat disfavored, but the
OLD model($1995$ physics) is perhaps 
still within the range of acceptability. Two
models are strongly disfavored(ruled out at a high significance
level). The no diffusion model,
NODIF, has a 
$0.45$\% rms difference in sound speed profilewith respect to the Sun;
this is 
$4.5$ times worse than the least
successful of the variant models. The model with a mixed solar
core, MIX(Cumming and Haxton 1996), has a rms difference of $1.8$\%, $18$
times worse than the least successful of the variant models. We
therefore conclude that the no diffusion model and the mixed model are
ruled out at a high level of significance.

\acknowledgments 
 
JNB is supported in part by NSF grant
\# PHY95-13835. 
MP is supported in part by NSF grant \# AST-9731621
We are grateful to Pawan Kumar for valuable discussions and
suggestions and to an anonymous referee for helpful comments on the
initial manuscript.
This work  utilizes data from the Solar Oscillations 
Investigation / Michelson Doppler Imager (SOI/MDI) on the Solar 
and Heliospheric Observatory (SOHO).  SOHO is a project of 
international cooperation between ESA and NASA. 
This work also utilizes data obtained by the Global Oscillation 
Network Group (GONG) project, managed by the National Solar Observatory, a 
Division of the National Optical Astronomy Observatories, which is 
operated by AURA, Inc. under a cooperative agreement with the 
National Science Foundation. The data were acquired by instruments 
operated by the Big Bear Solar Observatory, High Altitude Observatory, 
Learmonth Solar Observatory, Udaipur Solar Observatory, Instituto de 
Astrofisico de Canarias, and Cerro Tololo Inter-American Observatory. 
We would like to thank the Birmingham Solar Oscillation Network (BiSON) and
the   LOWL instrument  group for permission to use their data.  
BiSON is funded by the UK
Particle Physics and Astronomy Research Council.

\vfill\eject 
 
\clearpage 

\begin{table}[!t]
\begin{minipage}{6.25in} 
\center 
\caption{Properties of the solar models used. The first five models
listed all use input physics that is within the currently acceptable
range. The last four models are all deficient in one or more aspects
of the important input physics. 
\label{tab:diffmodels}} 
\vskip 1cm 
\begin{tabular}{llllll} 
\hline 
 
Model & $T_c$ & $\rho_c$ &  
      $Y_s$& $r_{\rm cz}/R_\odot$ & Comments \\ 
      &$10^6$ K & g cm$^{-3}$ & & & \\ 
 
\hline 
 
STD   & 15.74 & 152.98 & 0.2453 & 0.7123 & Standard model, incorporates 
diffusion, \\ 
      &       &        &        &        &only main sequence evolution \\ 
PMS   & 15.72 & 152.73 & 0.2455 & 0.7127 & Same as STD, but with pre-main 
sequence evolution \\ 
ROT   & 15.69 & 148.30 & 0.2530 & 0.7155 & Same as STD, but with 
rotational mixing of elements \\ 
R78   & 15.73 & 152.97 & 0.2454 & 0.7122 & Same as STD but with radius of 
695.78 Mm \\ 
R508  & 15.73 & 152.97 & 0.2454 & 0.7121 & Same as STD but with radius of 
695.508 Mm \\ 
NODIF & 15.44 & 148.35 & 0.2653 & 0.7261 & Same as STD but with no diffusion 
\\ 
OLD   & 15.80 & 154.52 & 0.2470 & 0.7111 & Old physics (see text) \\ 
S$_{34} = 0$ & 15.62 & 153.50 & 0.2422 & 0.7151 & Same as STD but with 
reaction constant 
S$_{34}$ set to 0 \\ 
MIX   & 15.19 &  90.68 & 0.2535 & 0.7314 & Core mixing (see 
text) \\ 
 
\hline 
\end{tabular} 
\end{minipage}
\end{table}

\clearpage

\begin{table}[!t]
\center
\caption{Solar sound speed, density, and adiabatic index, $\Gamma_1$, 
as derived from MDI data with model STD
\label{tab:solardata}}
\vskip 1cm
\begin{tabular}{ccccccc}
\hline
\noalign{\smallskip}
$r/R_\odot$& $c$& $\sigma_c$& $\rho$& $\sigma_\rho$ 
& $\Gamma_1$&  $\sigma_{\Gamma_1}$\\
       & (cm s$^{-1}$)&  (cm s$^{-1}$)&  (g cm$^{-3}$)
& (g cm$^{-3}$)\\
\noalign{\smallskip}
\hline
\noalign{\smallskip}
6.72421E-02& 5.11757E+07& 3.94099E+04& 1.15099E+02& 7.07543E$-$01&  1.66474E+00 & 2.25238E$-$03  \\
8.09010E-02& 5.11011E+07& 2.73209E+04& 1.03410E+02& 5.79980E$-$01&  1.66516E+00 & 1.91496E$-$03  \\
1.03240E-01& 5.06842E+07& 2.24633E+04& 8.56885E+01& 3.98898E$-$01&  1.66546E+00 & 1.52223E$-$03  \\
1.27841E-01& 4.97924E+07& 1.58364E+04& 6.89479E+01& 2.37507E$-$01&  1.66557E+00 & 1.11374E$-$03  \\
1.52136E-01& 4.85481E+07& 9.44176E+03& 5.52090E+01& 1.32744E$-$01&  1.66559E+00 & 7.36144E$-$04  \\
1.76272E-01& 4.70638E+07& 6.85910E+03& 4.39294E+01& 7.62635E$-$02&  1.66551E+00 & 5.23388E$-$04  \\
2.01621E-01& 4.53526E+07& 6.76634E+03& 3.42422E+01& 5.14520E$-$02&  1.66542E+00 & 6.15979E$-$04  \\
2.27095E-01& 4.36030E+07& 5.11622E+03& 2.63950E+01& 4.17573E$-$02&  1.66542E+00 & 8.55765E$-$04  \\
2.52021E-01& 4.19506E+07& 4.01399E+03& 2.02715E+01& 3.52018E$-$02&  1.66551E+00 & 1.06521E$-$03  \\
2.90019E-01& 3.95792E+07& 3.31986E+03& 1.33762E+01& 2.67010E$-$02&  1.66569E+00 & 1.24912E$-$03  \\
3.27903E-01& 3.74270E+07& 2.79850E+03& 8.76112E+00& 1.94585E$-$02&  1.66593E+00 & 1.30605E$-$03  \\
3.65873E-01& 3.54922E+07& 2.38388E+03& 5.72924E+00& 1.34765E$-$02&  1.66609E+00 & 1.32057E$-$03  \\
4.03759E-01& 3.37380E+07& 2.05408E+03& 3.77078E+00& 9.05253E$-$03&  1.66590E+00 & 1.30249E$-$03  \\
4.41725E-01& 3.21328E+07& 1.92635E+03& 2.50272E+00& 5.97644E$-$03&  1.66573E+00 & 1.26167E$-$03  \\
4.79653E-01& 3.06602E+07& 1.65346E+03& 1.68048E+00& 3.92439E$-$03&  1.66558E+00 & 1.20028E$-$03  \\
5.17591E-01& 2.92897E+07& 1.47936E+03& 1.14214E+00& 2.59164E$-$03&  1.66570E+00 & 1.12008E$-$03  \\
5.55534E-01& 2.79934E+07& 1.48527E+03& 7.85515E$-$01& 1.72961E$-$03&  1.66571E+00 & 1.02829E$-$03  \\
5.93468E-01& 2.67442E+07& 1.41520E+03& 5.46543E$-$01& 1.17043E$-$03&  1.66564E+00 & 9.32667E$-$04  \\
6.31376E-01& 2.55149E+07& 1.30023E+03& 3.84837E$-$01& 8.05904E$-$04&  1.66544E+00 & 8.38210E$-$04  \\
6.69240E-01& 2.42184E+07& 1.27796E+03& 2.74970E$-$01& 5.65616E$-$04&  1.66531E+00 & 7.46384E$-$04  \\
7.06965E-01& 2.26716E+07& 1.40422E+03& 2.00821E$-$01& 4.06955E$-$04&  1.66517E+00 & 6.53303E$-$04  \\
7.19541E-01& 2.20133E+07& 1.54541E+03& 1.82675E$-$01& 3.68780E$-$04&  1.66516E+00 & 6.22255E$-$04  \\
7.32138E-01& 2.13270E+07& 1.59282E+03& 1.66240E$-$01& 3.34657E$-$04&  1.66523E+00 & 5.90435E$-$04  \\
7.44804E-01& 2.06369E+07& 1.51408E+03& 1.50636E$-$01& 3.03069E$-$04&  1.66529E+00 & 5.59747E$-$04  \\
7.57440E-01& 1.99494E+07& 1.60979E+03& 1.35981E$-$01& 2.73225E$-$04&  1.66533E+00 & 5.29440E$-$04  \\
7.70062E-01& 1.92624E+07& 1.54903E+03& 1.22338E$-$01& 2.45603E$-$04&  1.66543E+00 & 4.99415E$-$04  \\
7.82731E-01& 1.85644E+07& 1.55268E+03& 1.09494E$-$01& 2.20125E$-$04&  1.66557E+00 & 4.71405E$-$04  \\
7.95369E-01& 1.78693E+07& 1.60056E+03& 9.75258E$-$02& 1.96246E$-$04&  1.66569E+00 & 4.45765E$-$04  \\
8.08017E-01& 1.71624E+07& 1.54036E+03& 8.63885E$-$02& 1.74422E$-$04&  1.66579E+00 & 4.20638E$-$04  \\
8.45936E-01& 1.49914E+07& 1.65047E+03& 5.74269E$-$02& 1.17902E$-$04&  1.66671E+00 & 3.61961E$-$04  \\
8.83841E-01& 1.26732E+07& 1.74022E+03& 3.46354E$-$02& 7.36665E$-$05&  1.66729E+00 & 3.37905E$-$04  \\
9.21650E-01& 1.00831E+07& 2.00392E+03& 1.74318E$-$02& 4.03160E$-$05&  1.66707E+00 & 3.66728E$-$04  \\
9.34217E-01& 9.11157E+06& 2.12779E+03& 1.28852E$-$02& 3.14810E$-$05&  1.66581E+00 & 3.84937E$-$04  \\
\noalign{\smallskip}
\hline
\end{tabular}
\end{table}

\clearpage
\begin{table}[!b]
\center
\begin{minipage}{3.00in}
\caption{The root-mean-squared (RMS)  sound speed, density, and
adiabatic index
differences with respect to the Sun.  The table lists the percentage
differences between the variables determined with the MDI data and
those predicted by different solar models considered in the
text (see Section~\ref{subsec:variants}).  
The first four rows refer to variant standard models; the last four
rows refer to different deficient (or non-standard) models.
This table shows how
different each of the models is from the Sun.
\label{tab:rms}}
\vskip 1 true cm
\begin{tabular}{@{\extracolsep{20pt}}llll}
\hline
\noalign{\smallskip}
Model &\multicolumn{1}{c}{$(\delta c/c)$} &
\multicolumn{1}{c}{$(\delta\rho/\rho)$}& \multicolumn{1}{c}{$(\delta \Gamma_1/\Gamma_1)$}\\
&\multicolumn{1}{c}{\%}&\multicolumn{1}{c}{\%}&\multicolumn{1}{c}{\%}\\
\noalign{\smallskip}
\hline
\noalign{\smallskip}
STD      & 0.069 & 0.942 & 0.087 \\
PMS      &  0.085 & 0.929 & 0.082\\
ROT      &  0.069 &  1.404 & 0.084 \\
R78     &  0.064 & 0.673 & 0.090 \\
R508    & 0.098 & 0.408 & 0.126 \\
OLD     & 0.170 & 0.905  & 0.126 \\
S$_{34}=0$   &  0.209 & 3.108 & 0.146 \\
NODIF    &  0.447 &  5.303 & 0.093\\
MIX     & 1.795 & 13.319 & 0.341\\
\noalign{\smallskip}
\hline
\end{tabular}
\end{minipage}
\end{table}

\begin{table}[!b]
\center
\begin{minipage}{4.0in}
\caption{Dependence upon reference model: This table
gives the rms differences in solar sound-speed,
 density, and adiabatic index obtained using different reference models
between radii of 0.07$R_\odot$ and $0.9R_\odot$. See the
introduction to Sect.~\ref{sec:dependence}
for the definition of how the
 reference model dependence is calculated.
\label{tab:tabdif}}
\vskip 1 true cm
\begin{tabular}{lccccc}
\hline
\noalign{\smallskip}
Model & \multicolumn{1}{c}{$({\Delta c\over c})$}  &
\multicolumn{1}{c}{$({\Delta c\over c})$}&\multicolumn{1}{c}{$({\Delta c\over 
c})$}&
\multicolumn{1}{c}{$({\Delta \rho\over \rho})$} 
& \multicolumn{1}{c}{$({\Delta \Gamma_1\over \Gamma_1})$}\\
\noalign{\smallskip}
 & MDI & BiSON+LOWL & GONG & MDI & MDI\\
&\multicolumn{1}{c}{\%}&\multicolumn{1}{c}{\%}&\multicolumn{1}{c}{\%}
&\multicolumn{1}{c}{\%} &\multicolumn{1}{c}{\%}\\
\noalign{\smallskip}
\hline
\noalign{\smallskip}
PMS        &  0.0246  & 0.0256 & 0.0131 &  0.2245 & 0.0124 \\
ROT        &  0.0179  & 0.0321 & 0.0373 &  0.3067 & 0.0251\\
R78        &  0.0253  & 0.0345 & 0.0415 &  0.1361 & 0.0150\\
R508       &  0.0685  & 0.0700 & 0.0781 &  0.3077 & 0.0476\\
OLD        &  0.0190  & 0.0250 & 0.0125 &  0.2206 & 0.1401 \\
S$_{34}=0$ &  0.0224  & 0.0251 & 0.0226 &  1.5556 & 0.0231\\
NODIF      &  0.0734  & 0.0465 & 0.0826 &  1.8858 & 0.0543\\
MIX        &  0.4112  & 0.5078 & 0.3948 &  3.3593 & 0.2899\\
\noalign{\smallskip}
\hline
\end{tabular}
\end{minipage}
\end{table}

\end{document}